\documentstyle[epsfig]{aipproc}

\newcommand{\etal}{\emph{et al.}}

\renewcommand{\vec}[1]{{\bf #1}}
\begin{document}

\title{Mean--Field Models\\ and Superheavy Elements}

\author{P.--G. Reinhard,$^{1,2}$
        M. Bender,$^3$ and
        J. A. Maruhn$^{2,4}$
}
\address{$^1$Institut f\"ur Theoretische Physik II, 
             Universit\"at Erlangen--N\"urnberg,\\ 
             Staudtstra{\ss}e 7, 
             D--91058 Erlangen, Germany\\[1mm]
         $^2$Joint Institute for Heavy-Ion Research, 
             Oak Ridge National Laboratory,\\
             Oak Ridge, TN 37831--6374, U.S.A.\\[1mm]
         $^3$Gesellschaft f\"ur Schwerionenforschung,\\ 
             Planckstra{\ss}e 1, 
             D--64921 Darmstadt, Germany\\[1mm]
         $^4$Institut f\"ur Theoretische  Physik, 
             Universit\"at Frankfurt am Main,\\
             Robert--Mayer--Stra{\ss}e 8--10, 
             D--60325 Frankfurt, Germany
}
\maketitle
%
%
\begin{abstract}
We discuss the performance of two widely used nuclear mean-field
models, the relativistic mean--field theory (RMF) and the
non-relativistic Skyrme--Hartree--Fock approach (SHF), with particular
emphasis on the description of superheavy elements (SHE).
We provide a short introduction to the SHF and RMF, the relations between
these two approaches and the relations to other nuclear structure
models, briefly review the basic properties with respect to normal
nuclear observables, and finally present and discuss recent results on 
the binding properties of SHE computed with a broad selection of 
SHF and RMF parametrisations. 
\end{abstract}
%
%
\section{Introduction}
Nuclear structure models are available at various levels of
description. There is the macroscopic view in terms of the
liquid-drop model (LDM) \cite{LDM}. The macroscopic-microscopic
(mac-mic) method combines the rich phenomenological experience
summarised in the LDM with a fine-tuning through shell effects
estimated in a properly chosen single-particle potential
\cite{micmac,micmac2}. And there is the broad family of self-consistent
mean-field approaches (Hartree or Hartree--Fock) employing effective
energy functionals on which we will concentrate here.  All these
models presently enjoy a revival due to a world of new experimental
information emerging from the production and measurement of exotic
nuclei and new elements. In fact, it is more than three decades ago
that speculations on the possible existence of shell-stabilized
superheavy elements (SHE) \cite{Mosel,SuperNils} have motivated the
construction of dedicated heavy-ion accelerators.  The production of
SHE turned out to be the most tedious task in the field of exotic
nuclei. It took about two decades to reach the first island of
shell-stabilised deformed SHE in the region of \mbox{$Z = 108$}
\cite{Dubna,GSI,GSI2}. Recent experiments give first evidence for
nuclei even closer to the expected island of spherical SHE.  The
synthesis of the neutron-rich isotopes \mbox{$^{283}112$},
\mbox{$^{287-289}114$} \cite{Z114}, and 
\mbox{$^{292}116$} \cite{Z116} were reported from Dubna and at
Berkeley three $\alpha$-decay chains attributed to the even heavier
${}^{293}118$ were observed \cite{Z118}.  While earlier superheavy
nuclei could be unambiguously identified by their $\alpha$-decay
chains leading to already known nuclei, the decay chains of the
new-found superheavy nuclei cannot be linked to any known nuclides.
The new discoveries still have to be viewed carefully, see the
critical discussion in \cite{Arm00a}.  While for the heaviest systems
only their mere existence and a few decay properties are established,
the first spectroscopic data become available for nuclei at the lower
end of the superheavy region, e.g.\ low-lying states in Rf isotopes
from the analysis of $\alpha$-decay fine-structure \cite{Hessberger}
and rotational bands of nuclei around $^{254}$No which were found to
be stable against fission at least up to \mbox{$I=16$} \cite{Reiter,Leino}.
Interpretation of data and planning of future experiments call for a
significant refinement in the modeling of SHE.  As their mere
existence emerges from a delicate balance between the Coulomb instability of
the liquid drop against fission and stabilisation through
shell effects, SHE provide a demanding testing ground for nuclear
structure models, probing all their details.  The aim of this review is to 
examine the performance of current nuclear mean-field models under the
particular perspective of SHE. In this context we concentrate on the two
most widely used brands, the relativistic mean-field model (RMF) and
the non-relativistic Skyrme-Hartree-Fock approach (SHF).  We ought to
mention that there are also other models like the non-relativistic
Gogny force \cite{gogny,gogny2} employing finite-range terms in the
interaction, the energy functionals of Fayans \etal\ \cite{fayans}
which are similar to SHF but use variants of density dependence and
pairing interaction, or the point-coupling variant of the RMF
\cite{madland} which can be viewed as the relativistic analogue of
the Skyrme interaction. As none of these was widely used for the
calculation of SHE so far we omit them from our discussion.

The paper is outlined as follows: 
Section \ref{sec:frame} provides the theoretical background of the
RMF and SHF, tries to establish the relation between these two
approaches and the relations to the more macroscopic methods
(LDM and mic-mac). 
Section \ref{sec:result} presents and discusses a selection of
typical results.
%
%
\section{Framework}
\label{sec:frame}
%
%
\subsection{Mean-field models in the hierarchy of approaches}
Self-consistent mean-field models are intermediate between the fully 
microscopic 
many-body theories as, e.g., Br\"uckner--Hartree--Fock (BHF) \cite{BHF} 
and semi-classical models as the mac-mic approach \cite{micmac,micmac2}.
The microscopic approaches have made considerable
progress over the past decades \cite{BHF,HenPan}, yet the actual
precision in describing nuclear properties is still limited. Moreover,
application to finite nuclei is extremely expensive. Thus fully
microscopic methods are presently not used for large-scale nuclear
structure calculations. They provide, however, useful guidelines 
for the construction of effective mean-field theories \cite{Tmatexp}.

On the other side are the macroscopic approaches which are inspired
by the idea that the nucleus is a drop of nuclear liquid, giving rise 
to the liquid-drop model (LDM) and the more refined droplet 
model \cite{droplet}. With a mix of intuition and systematic expansion 
one can write down the corresponding energy functional even
including finite-range effects of the nuclear interaction \cite{FRDM,FY}. 
There remain a good handful of free parameters, as 
e.g.\ the coefficients for volume energy \mbox{$a_{\rm vol}=E/A$}, 
symmetry energy $a_{\rm sym}$, 
incompressibility $K_\infty$, or surface energy $a_{\rm surf}$. 
These have to be adjusted to a multitude of nuclear bulk
properties such that modern droplet parametrisations deliver 
an excellent description of average trends \cite{LDM}.
Actual nuclei, however, deviate from the average due to 
quantum shell effects, so that shell corrections are added,
which are related to the level density near the Fermi surface and
can be computed from a well tuned nuclear single-particle 
potential. Macroscopic energy plus shell corrections constitute the 
mac-mic approach which is enormously successful in reproducing the 
systematics of known nuclear binding energies \cite{micmac,micmac2,FRDM,FY}. 
One has to admit, though, that the mac-mic method relies
strongly on phenomenological input. This induces uncertainties
when extrapolating to exotic nuclei. Particularly uncertain is the
extrapolation of the single-particle potential because this is not
determined self-consistently but added as an independent piece of
information.

Self-consistent mean-field models do one big step towards a
microscopic description of nuclei. They produce the appropriate 
single-particle potential corresponding to the actual density 
distribution for a given nucleus. Still, they cannot be handled 
as an ab initio treatment because the genuine nuclear interaction 
induces huge short-range correlations. Self-consistent 
mean-field models deal with effective energy functionals. 
The concept has much in common with the
successful density-functional theory for electronic systems
\cite{GD90,Petkov}. Taking up the notion from there, we can speak of nuclear
Kohn-Sham models as synonym for mean-field models. The difference is,
however, that electronic correlations are well under control and that
reliable electronic energy-density functionals can be derived ab
initio. Nuclear many-body theories, as discussed above, have not yet
reached sufficient descriptive power to serve as immediate input for
effective mean-field models, but serve as motivation and source
for the basic features of the mean-field approach. This sets the
framework, the actual energy functionals are then constructed by
systematic expansion considering symmetries \cite{Dob96a}, their 
parameters are adjusted phenomenologically.
(For a recent review on nuclear correlations and their relation to
effective mean-field theories see \cite{Rei94a}.)

The connection between self-consistent mean-field models and mac-mic
approaches is much better developed. There are several attempts from
either side. The ETFSI approach starts from SHF and derives an
effective mac-mic model by virtue of a semi-classical expansion
\cite{ETFSI}. From the macroscopic side there is an attempt to 
induce more self-consistency by virtue of a Thomas-Fermi approach
\cite{TFmyers}. The investigation of these links is useful to gain
more insight into the crucial constituents of either models.  
%
%
\subsection{Skyrme--Hartree--Fock model} 
The concept of an effective interaction for mean-field calculations 
can be justified within
many-body theories. For example, the $T$-matrix in BHF is the actual
effective force for the underlying mean field. It was a formal
$T$-matrix expansion \cite{Tmatexp} which gave theoretical support to 
the first working self-consistent model \cite{Skfirst} using an
effective interaction introduced much earlier by T.~H.~R.\ Skyrme 
\cite{skyrme,skyrmeLS}. 
The original idea was that a convenient-to-use effective interaction 
can be obtained from a momentum-space expansion of any finite-range
interaction which leads to a zero-range force plus momentum-dependent
terms. A density dependence has to be added to incorporate many-body 
correlations in an effective way (note that the $T$ matrix depends 
strongly on density) and, last not least, a (zero-range) spin-orbit 
force \cite{skyrmeLS} is added to account for the strong spin-orbit
splitting in nuclei.

This concept has an intimate relation to energy-density functionals. 
The energy expectation value of such an effective interaction is precisely 
a functional of the local density. Thus the well developed
density-functional theory \cite{GD90} adds support for SHF. 
We present here the SHF functional complemented by an obvious 
graphical illustration:
\begin{figure}[h!]
\centerline{\epsfig{figure=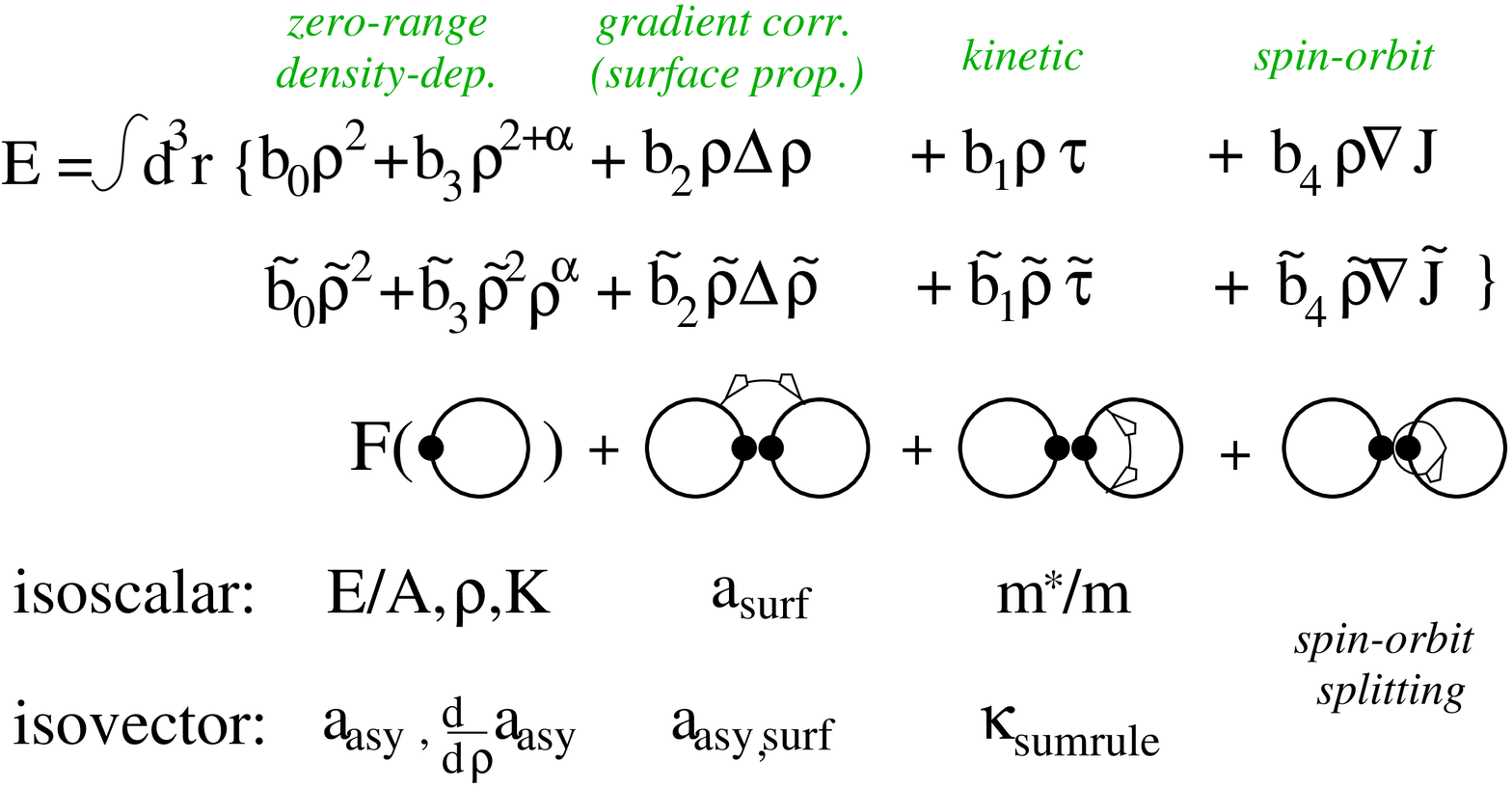,height=12.5cm,angle=-90}}
\end{figure}
The first
two columns parametrise a density functional marked $F(\rho)$.
Note the distinction between total (isoscalar) density 
\mbox{$\rho=\rho_{\rm p}+\rho_{\rm n}$} and isovector density 
\mbox{$\tilde{\rho}=\rho_{\rm p}-\rho_{\rm n}$}, and similarly 
for the kinetic density $\tau$ and the spin-orbit current $\vec{J}$.
The leading term is the two-body interaction \mbox{$\propto \rho^2$}. It is
attractive in the isoscalar channel and repulsive for the isovector
part, just as the genuine two-body force is. The necessary 
density-dependent interaction is parametrised in the next term, for
traditional reason in the form of an extra $\rho^\alpha$. These two
terms together set up a local-density functional which is 
represented graphically as $F(\rho)$ where a heavy dot with an 
appended circle stands for the density. That density functional
allows already to fix all relevant nuclear bulk properties. 
Finite systems require a fine-tuning of surface properties
which is achieved by the gradient correction term. They are
represented graphically by the right-left arrow atop the $\rho*\rho$ symbol.
Moreover, the strong dressing of nucleons in matter calls for an 
effective mass \mbox{$m^*/m < 1$} \cite{effmass}
which can be achieved by the kinetic correction term
$\propto\rho\tau$ where \mbox{$\tau=\sum_\alpha|\nabla\varphi_\alpha|^2$}. 
This term requires derivatives within the density summation which is
indicated by an up-down arrow in one of the two densities.
Last but not least, a strong spin-orbit splitting is
crucial for a correct description of single-particle spectra
and shell closures \cite{lsfirst}. This is guaranteed by 
the spin-orbit term \mbox{$\propto\rho\nabla\cdot \vec{J}$} 
where the spin-orbit current is indicated graphically by a circle
with arrow around the density-dot. Note that each term comes twice, once 
in an isoscalar form and another time in an analogous isovector form.

The nice feature of the Skyrme functional is that each term can be
related to a corresponding bulk property (or LDM feature). This is
indicated in the last two lines of the above sketch. The two isoscalar
density-dependent parts together are related to bulk binding $E/A$,
equilibrium density $\rho_0$, and incompressibility $K_\infty$. 
The isovector part complements this by the symmetry-energy coefficient 
$a_{\rm sym}$ and its derivative \mbox{$\partial_\rho \, a_{\rm sym}$}. 
The gradient corrections relate naturally to the isoscalar and isovector 
surface-energy coefficients. The kinetic terms adjust the isoscalar 
effective mass $m^*/m$ and the isovector effective mass $m'{}^*$. The 
latter modifies the Thomas-Reiche-Kuhn sum rule \cite{RingSchuck}
by an enhancement factor \mbox{$1+\kappa_{\rm sumrule} = m/m'{}^*$} and 
one often parametrises $m'{}^*$ in terms of this enhancement factor, see 
e.g.\ \cite{SLyx}. No bulk
property can be associated directly with the spin-orbit term. This term is
related to the shell structure, i.e.\ to the single-particle spectrum,
of finite nuclei. What looks here like a quickly drawn and superficial
analogy to the LDM, has indeed deep theoretical foundations. One can, 
in fact, derive the mic-mac method from SHF by virtue of semi-classical
expansions \cite{SHF2micmac}.

There is a subtle difference between the force concept and the energy
functional concept which concerns the spin-orbit force. Derivation of
the energy functional from a Skyrme force yields (usually small) 
extra terms $\vec{J}^2$ and $\tilde\vec{J}^2$ which emerge from the 
exchange part of the kinetic terms $\propto\rho\tau$ or 
$\tilde{\rho}\tilde{\tau}$. Some Skyrme parametrisations include these 
terms, some do not. We will specify that later when presenting the 
parametrisations. The actual Hartree-Fock (or Kohn-Sham) equations are 
derived variationally from the given energy functional,
see e.g.\ \cite{SkIx,RPAnucl}.
%
%
\subsection{Relativistic mean-field model}
The history of the RMF has similarities to SHF. After an early first
conjecture \cite{duerr}, it was only in the seventies that this model
was lifted to a competitive mean-field model \cite{RMF1,RMF2}. The
starting point is, at first glance, different from SHF. RMF is
conceived as a relativistic theory of interacting nucleonic and
mesonic fields. The mesonic fields are approximated to mean fields
(real-number fields rather than field operators), a feature which is
reflected in the name RMF. Moreover, the anti-particle contributions
in the Dirac fields for the nucleons are suppressed (``no--sea''
approximation). Again, the mean-field approximation is not valid in
connection with the true physical meson fields. The meson fields of
the RMF are effective fields at the same level as the forces in SHF
are effective forces. The RMF is the relativistic cousin of SHF,
and the same strategy applies: relativistic BHF is still not precise
enough to allow an ab initio derivation of the RMF; the model is
postulated from a mix of intuition and theoretical guidelines with the
parameters to be fixed phenomenologically.

The RMF is usually formulated in terms of an effective Lagrangian, as
any relativistic theory. For stationary problems it can be mapped to an 
energy functional \cite{Schmid}. We discuss it here in terms of an 
effective energy functional and we confine ourselves to a graphical
presentation because the RMF is well documented in several reviews
\cite{Rei89,Ser92a,Rin96a}. The RMF functional can be sketched as follows:
\begin{figure}[h!]
\centerline{\epsfig{figure=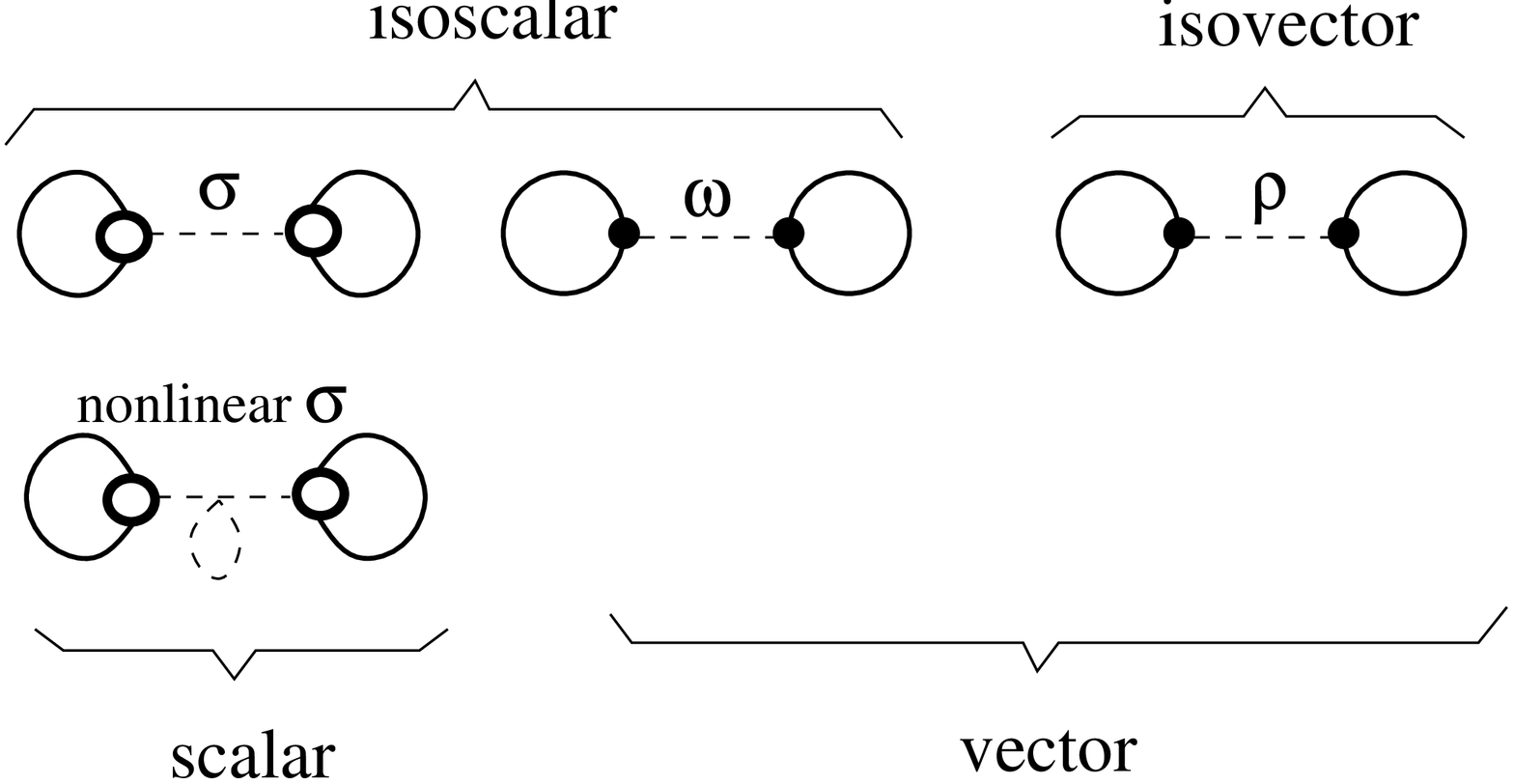,height=9.5cm,angle=-90}}
\end{figure}
The \emph{ansatz} looks at first glance conceptually even simpler than that
of SHF. One merely writes down the basic nucleon-meson couplings
\cite{RMF1}. The mesons can be characterised by their internal quantum
numbers. Scalar and vector fields are taken into account (the most
famous pion field does not contribute in Hartree approximation because 
there is no finite pseudo-scalar density in the ground state). 
For the isovector part, one employs only the vector field 
associated with the $\rho$ meson. The
isovector-scalar field, the $\delta$ meson, should appear at the same
level (and plays a role, indeed, in the nucleon-nucleon force). It can
be omitted for purely phenomenological reasons because it does not
improve the performance of the model when included. But a simple 
series of meson-nucleon couplings does
not suffice to deliver a high-precision model. We know from
microscopic theory that the effective interaction needs density
dependence to effectively incorporate many-body correlations. 
Such had been introduced into the RMF via non-linear terms
(cubic and quartic) in the scalar meson field \cite{RMF2}, as
indicated by the second line in the above sketch. This leads to
a model with the same descriptive power as SHF \cite{Rei89,Ser92a,Rin96a}.
This choice to introduce density-dependence was originally motivated
by the aim to maintain renormalisibility of the theory. This is not
a very stringent condition in connection with an effective mean-field
theory which incorporates many-body effects, but the ansatz delivers
an empirically well-working scheme and thus there was little pressure
for modifications, for exceptions and variants of the modeling see
\cite{PL40,TM1}.

At first glance, the RMF functional looks quite different from SHF, and
indeed, the pieces have been put together in a different fashion.  The
concept of the RMF seems to emerge naturally from a field theoretical
perspective whereas the many-body aspects are not so
transparent. These had been more obvious in SHF, particularly the
relation to nuclear bulk properties, but it is possible to draw
straight connections from RMF to SHF.  These can be established by
considering the non-relativistic and zero--range limit of RMF 
\cite{Rei89,thies}. It can be sketched as follows:
\begin{figure}[h!]
\centerline{\epsfig{figure=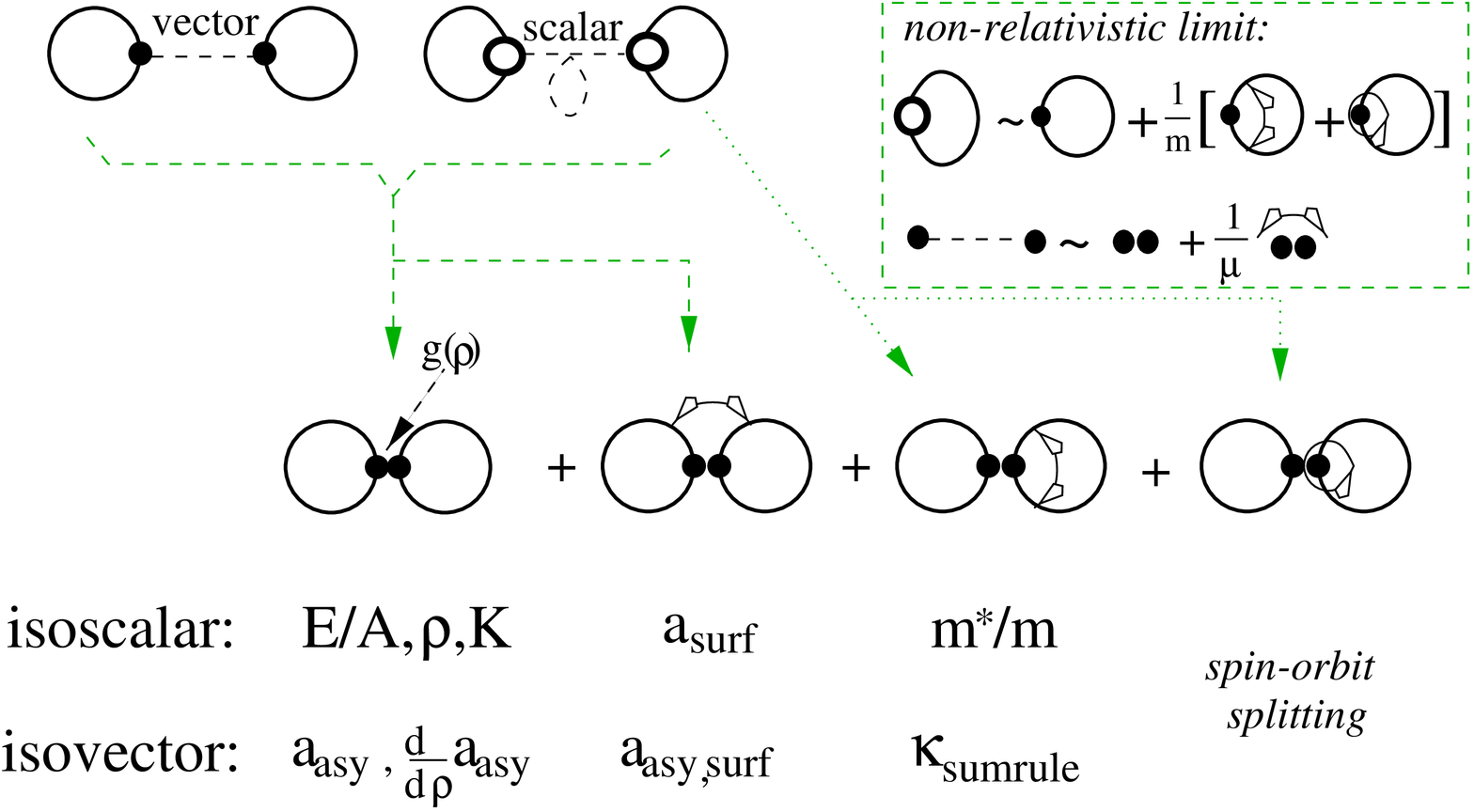,height=12.5cm,angle=-90}}
\end{figure}
The left upper two diagrams summarise the RMF from the previous
sketch. The right upper box indicates the two independent steps of
expansion: a $v/c$ expansion of the scalar density and a gradient
expansion of the meson propagator.  The scalar density delivers as
leading term the normal density (zero component of vector density) and
as $v/c$ corrections the kinetic-energy density $\tau$ as well as the
divergence of the spin-orbit current $\nabla \cdot \vec{J}$. 
The finite range of the mesons is expanded as leading zero-range 
coupling and gradient correction. 
Inserting these approximations yields the functional as
represented by the diagrams in the second line of the sketch. It looks
almost identical to the Skyrme functional. The interpretation in terms
of bulk properties then proceeds as in case of SHF.  It is repeated
here for the sake of completeness. There is one aspect, however, which
is very hard to map: it is the form of the density dependence. The
mechanism in the RMF goes through non-linear meson coupling and is
much different from the SHF with its straightforward expansion in
powers of density $\rho$. A thorough comparison of the density
dependences is still a task for future research.
As in case of the SHF, we skip a detailed derivation of the coupled
field equations and refer the reader to \cite{Rei89,Ser92a,Rin96a}.
%
%
\subsection{Further ingredients}
\label{sec:ingred}
The above two subsections have outlined the main body of SHF and RMF.
There are several further details which are handled similarly in both
approaches.
The direct part of the Coulomb interaction is given by the standard
expression
$
E_{\rm coul}
= \frac{1}{2} \int \! d^3r \, d^3r' \; \rho_c(\vec{r}) \, 
  |\vec{r}-\vec{r}'|^{-1} \rho_c(\vec{r}')
$ 
where the charge density $\rho_c$ is usually replaced by the mere
proton density \mbox{$\rho_c \rightarrow \rho_p$} omitting
substructure and finite size of the nucleons.
While the RMF includes the direct term only, all modern SHF 
parameterizations employ the Slater approximation for the
exchange term
\mbox{$
E_{\rm coul,ex} 
= \frac{3}{4} e^2 \int \! d^3r \, \rho_p^{4/3} (\vec{r})$}
\cite{GD90}. 

A further crucial ingredient are pairing correlations. 
There are several recipes in the literature differing in the
variational principle used, the correction for the particle-number 
uncertainty of the BCS state and the effective pairing interaction. 
Nowadays, for the latter a zero-range two-body pairing force 
\mbox{$V_{\rm pair} = V_{0,p/n} \delta(\vec{r}_1-\vec{r}_2)$} 
with separately adjustable strengths for protons and neutrons
is most widely used. For most calculations reported here 
the matrix elements of this force are used in the BCS equations 
(as approximation to Hartree--Fock--Bogoliubov).
For a recent discussion of this and competing recipes see 
\cite{gappaper} and references therein.

The mean field localises the nucleus in space. This violates
translation invariance. Center-of-mass projection restores that
symmetry \cite{RingSchuck}. It turns out that a second-order estimate 
for the center-of-mass correction is fully sufficient. The correction 
is performed by subtracting
\mbox{$E_{\rm cm}^{\rm(2)} = \langle\hat{P}_{\rm cm}^2\rangle/2mA$}
from the calculated binding energy where $m$ is the nucleon mass 
and $A$ the mass number, see \cite{cmpaper} for details. The term is 
usually subtracted a posteriori to circumvent two-body terms in the 
mean-field equation. For some parameterizations this is even simplified 
further. One approach is to use the harmonic oscillator estimate
\mbox{$E_{\rm cm}^{\rm (est)} = \frac{3}{4} 41.5 \, {\rm MeV} A^{-1/3}$},
another to use only the diagonal part of $\hat{P}_{\rm cm}^2$,
i.e.\ \mbox{$E_{\rm cm}^{\rm(diag)}=\sum_i \langle\hat{p}_i^2 \rangle / 2mA$}.
The latter recipe leads to a simple renormalisation of the nucleon
mass \mbox{$1/m\rightarrow 1/m \times (1-1/A)$} and is usually included in the
variational equations. Different groups are using different recipes
for the center-of-mass correction and thus one has to keep 
track which recipe is employed with a given parametrisation.

In fact, center-of-mass correction is already one step, although the
most trivial, beyond the mean-field approach. 
There are many other correlation effects possibly to be considered, 
particle-number projection,
angular-momentum projection in case of deformed nuclei, vibrational
corrections. These aspects are being investigated intensively at present. 
Here we stay at the strict mean-field level (plus c.m.\ correction).
%
%
\subsection{Parametrisations}
As already mentioned above, nuclear many-body theory is not yet precise enough
to allow an ab initio derivation of the effective energy functionals
for SHF or RMF from nucleon-nucleon interactions. Theory, with a spark 
of intuition, sets the frame and defines the form of the functional. 
The remaining free parameters have to be adjusted phenomenologically. 
Different groups have different biases in selecting the observables to 
which a force should be fitted. One usually restricts the fits to a few 
spherical nuclei with at least one magic nucleon number
(an exception is a recent large-scale fit to all known nuclear masses
\cite{pearstond}). All fits take care of binding energy $E_B$ and
r.m.s.\ charge radii $r$. From then on, different tracks are pursued.
SHF fits invoke extra information on spin-orbit splittings. RMF
generally does not need that because the spin-orbit interaction
is a relativistic effect that emerges naturally from relativistic
models. Some groups add information on nuclear matter, some even on 
neutron matter. Some groups give a weight to
isovector trends. Others make a point to include more information from
the electromagnetic formfactor, in terms of a diffraction radius and
surface thickness \cite{Fri86a}. A detailed discussion of a fitting
strategy can be found, e.g., in \cite{SkIx,Rei89,Fri86a}.

In view of these different prejudices entering the determination of a
force, it is no surprise that there exists a world of different
parametrisations for SHF as well as RMF. We confine the discussion to
a few well adjusted and typical sets. For SHF we consider the
parametrisations ${\rm SkM}^{*}$ \cite{SkM*}, SkP \cite{SkP}, SkT6
\cite{Tx}, Z$_\sigma$ \cite{Fri86a}, SkI3, SkI4 \cite{SkIx}, and SLy6
\cite{SLyx}.  The forces SkM$^*$, SkT6, SkP, and Z$_\sigma$ can be
called the second generation forces which emerged in the mid eighties and
which delivered for the first time a well equilibrated high-precision
description of nuclear ground states.  The force ${\rm SkM}^{*}$ was
the first to deliver acceptable incompressibility and fission
properties.  It also provides a fairly good description of surface
thickness although this type of data was not fitted explicitly.  The force
SkT6 is a fit with constraint on \mbox{$m^*/m = 1$}. It did take
into account the nuclear surface energy and thus also provides a satisfying
surface thickness ($\equiv$ electromagnetic formfactor).  The force
SkP uses effective mass \mbox{$m^*/m = 1$} and is designed to allow a
self-consistent treatment of pairing.  We will skip this pairing
feature and use an appropriately adjusted delta pairing force.
The force Z$_\sigma$ stems from a least-squares fit including
diffraction radius and surface thickness but without any reference to
pseudo-data from nuclear matter.  The forces SLy6, SkI3, and SkI4 have
been developed in the nineties.  They take care of new data (e.g.\ from
exotic nuclei) and new aspects.  The force SLy6 stems from a recent
attempt to cover properties of pure neutron matter together with
normal nuclear ground state properties, sacrificing the
quality of surface thickness somewhat to achieve this. All Skyrme forces up to
here use the spin-orbit coupling in the particular combination
\mbox{$3\rho \nabla \cdot \vec{J}+\tilde{\rho} \nabla \cdot \tilde\vec{J}$}
which is dictated by deriving the spin-orbit energy from a two-body 
zero-range spin-orbit force \cite{skyrmeLS}. The forces SkI3/4 employ a
spin-orbit force with isovector freedom to simulate the relativistic
spin-orbit structure. SkI3 contains a fixed isovector part
\mbox{$\tilde{b}_4=0$} analogous to the RMF, whereas SkI4 is adjusted allow
free variation \mbox{$b_4 \neq 3 \tilde{b}_4$} of the isovector spin-orbit
force. The modified spin-orbit force was introduced because no
conventional SHF force was able to reproduce the isotope shifts of the
m.s.\ radii in heavy Pb isotopes, see \cite{SkIx} and references
cited therein. The isovector-modified spin-orbit force in SkI3 and
SkI4 solves this problem. It then has, of course, a strong effect
on the spectral distribution in heavy nuclei and thus for the
predictions of SHE.

For the RMF we consider the parametrisations NL--Z \cite{NLZ},
NL3 \cite{NL3} and TM1 \cite{TM1}.
The force NL--Z comes from fits with the choice of observables quite similar 
to those of SkI3 and SkI4, with in particular the charge formfactor 
taken care of. NL3 is fitted without looking at the formfactor but
more emphasis on the isovector trends.
TM1 is  an extended version of the RMF including a quartic
non-linear self-coupling of the isoscalar-vector field.

Each parametrisation is complemented by Coulomb, pairing and
center-of-mass correction as outlined in section~\ref{sec:ingred}. The
pairing strengths need to be adjusted separately to comply with the
level density of the force. Table~\ref{tab:pair} provides the
actually used pairing strengths. The table indicates also 
the type of center-of-mass correction used. 
%
%
\begin{table}[t!]
\caption{\label{tab:pair}
Proton and neutron pairings strengths (in [MeV fm$^3$]) and center-of-mass
recipe for the parametrisations used in this paper. For details of the
adjustment of the pairing strengths and the cutoff used see
\protect\cite{gappaper}.
}
\begin{tabular}{ccccccccccc}
 & SkM$^*$ & SkT6 & SkP & Z$_\sigma$ & SLy6 & SkI3 & SkI4 & NL-Z & NL3 & TM1 \\
\noalign{\smallskip}\hline\noalign{\smallskip}
$V_{0,p}$ & $-292$ 
          & $-256$
          & $-265$
          & $-290$
          & $-320$
          & $-350$
          & $-323$
          & $-351$
          & $-342$
          & $-327$ \\
$V_{0,n}$ & $-276$ 
          & $-250$
          & $-241$
          & $-269$
          & $-308$
          & $-340$
          & $-310$
          & $-349$
          & $-329$
          & $-323$ \\
\noalign{\smallskip}\hline\noalign{\smallskip}
c.m. & $E_{\rm cm}^{\rm(diag)}$
     & $E_{\rm cm}^{\rm(diag)}$
     & $E_{\rm cm}^{\rm(diag)}$
     & $E_{\rm cm}^{\rm(2)}$
     & $E_{\rm cm}^{\rm(2)}$
     & $E_{\rm cm}^{\rm(2)}$
     & $E_{\rm cm}^{\rm(2)}$
     & $E_{\rm cm}^{\rm(2)}$
     & $E_{\rm cm}^{\rm(est)}$
     & $E_{\rm cm}^{\rm(est)}$ \\
\noalign{\smallskip} 
\end{tabular}
\end{table}
%
%
\section{Results and discussion}
\label{sec:result}
%
%
\subsection{Basic properties}

\begin{figure}[bht]
\begin{center}
\centerline{\epsfig{figure=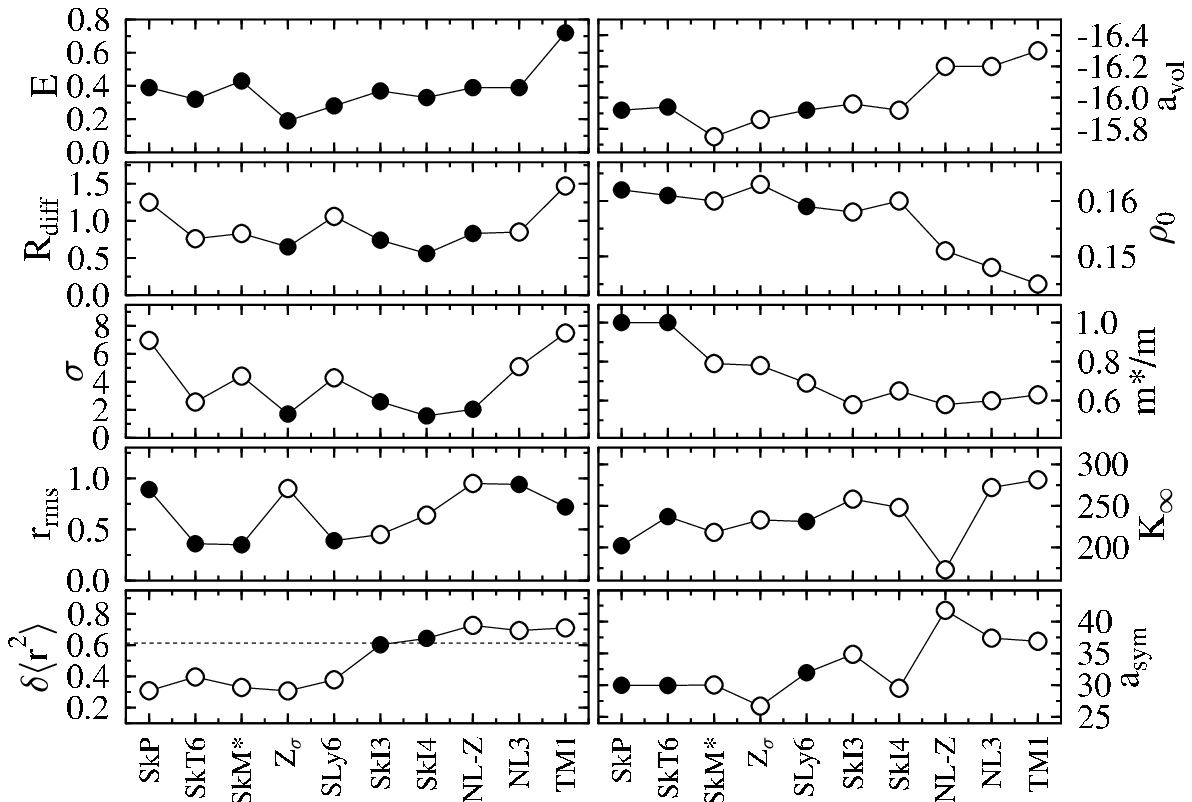}}
\end{center}
\caption{\label{fig:quality}
Left panels: relative errors (in $\%$) of key observables
of finite nuclei for the selection of forces:
binding energy $E$, 
diffraction radius $R_{\rm diff}$, 
surface thickness $\sigma$, and 
r.m.s.\ radius $r_{\rm rms}$.
The lowest left panel shows the isotopic shift of the m.s.\ 
radius in $^{214}$Pb,
\mbox{$\delta\langle r^2\rangle=r^2(^{214}{\rm Pb})-r^2(^{208}{\rm
Pb})$}, in units of ${\rm fm}^2$.
Its experimental value is indicated by a horizontal dotted line.
Right panels: properties of the model system of infinite 
homogeneous symmetric spin-saturated nuclear matter: 
volume-energy coefficient $a_{\rm vol}$ (in MeV), 
equilibrium density $\rho_0$ (in fm$^{-3}$), 
(isoscalar) effective mass $m^*/m$ (dimensionless),
incompressibility $K_\infty$ (in MeV), and 
symmetry-energy coefficient $a_{\rm sym}$ (in MeV).
Full symbols denote quantities that were used in the fit of the
particular effective interaction, while open symbols represent 
predictions.
}
\end{figure}
Before coming to a discussion of SHE we review briefly
the basic properties of the various parametrisations, their
performance with respect to normal nuclei and their nuclear matter
properties (which are equivalent to the coefficients of the LDM
expansion). They are summarised in Fig.~\ref{fig:quality}. 
Note that only a sub-set of these data are used in actual fits. 
The left
panels deal with finite nuclei. They show the r.m.s.\ errors of the
basic ground-state properties (relative errors in $\%$). Note that
diffraction radius and surface thickness are key quantities
determining the electromagnetic formfactor \cite{Fri86a}. For the
nuclei discussed here $R_{\rm diff}$, $\sigma$ and $r_{\rm rms}$
are linked by the Helm model in such a way that only two values are
independent \cite{FriVoe}. The lowest
panel adds a more specific piece of information: the isotopic shift in
heavy Pb isotopes. The energy (uppermost panel) is very well
reproduced. All chosen forces have an error of only $0.4\%$ or below.
More differences are seen concerning the reproduction of radii and
surface thicknesses. The correlation is obvious: quantities which had
been included in the adjustment (full dots) are usually well
reproduced. Those which had not been fitted tend to show larger errors.
Exceptions from the rule to some extent are SkT6 and SkM$^*$ which
yield acceptable $R_{\rm diff}$ and $\sigma$ without having fitted
them. Both forces, however, include a fitted surface energy which is
related to reasonable surface thickness \cite{Fri86a}. A less positive
exception is the comparatively large error in $r_{\rm rms}$ for NL3, which
includes this observable in the fit, but the other RMF forces have
similar problems with $r_{\rm rms}$. It seems that this is a principal
problem of the RMF in its present form. It could be related to the
somewhat curious form of shaping the density dependence in that
approach. After all, one can conclude that the error in radii, 
$R_{\rm diff}$ or $r_{\rm rms}$, 
is certainly below $1\%$, often half of that. Surface thicknesses
$\sigma$ can be reproduced within $2\%$ if used in the fit; 
moreover, $\sigma$ gives a handle on the surface tension (and
subsequently on good fission barriers \cite{SkM*,cmpaper}).
The lowest right panel in Fig.~\ref{fig:quality} shows the isotopic
shift in heavy Pb isotopes. It is obvious that all 
conventional Skyrme forces (i.e.\ those with \mbox{$b_4=\tilde{b}_4$}) 
fall short of the experimental value of
\mbox{$\delta\langle r^2\rangle=r^2(^{214}{\rm Pb})-r^2(^{208}{\rm Pb})=0.6\,
{\rm fm}^2$}. 
All RMF forces hit that value very well as a prediction.
It was worked out that this is due to the particular form of the
spin-orbit force in the RMF \cite{SkIx,Ringls}. Extending the
SHF to allow for \mbox{$b_4\neq\tilde{b}_4$} yields an equally good
reproduction of these isotopic shifts, see SkI3 and SkI4 in
Fig.~\ref{fig:quality}. But the values need to be included as fit
data because the spin-orbit force is added ``by hand'' in SHF whereas
it is an intrinsic feature of the nucleonic Dirac equation in RMF.

The right panels of Fig.~\ref{fig:quality} show nuclear matter
properties. There is general agreement about the volume energy,
although the RMF forces seem to prefer slightly smaller values.
The equilibrium density is almost the same for all SHF forces while
RMF again prefers slightly smaller values. This systematic difference
in extrapolation to nuclear matter is most probably related to the
very different way in which the density-dependence is modeled in SHF
and RMF. A thorough study of those effects is still lacking.

The effective mass shows a clear trend to values lower than one.  It
is, however, a rather vaguely fixed property. For example, SkT6 has
fixed \mbox{$m^*/m=1$} and is still able to provide good overall quality (see
right panels). It is said that fits which concentrate on binding
energies automatically prefer \mbox{$m^*/m=1$} \cite{pearstond}. On the other
hand, fits which include the formfactor ($R_{\rm diff}$ and $\sigma$)
prefer lower $m^*/m$. And the RMF always prefers particularly low values.
It is yet an open point what the best value for $m^*/m$ should be for
nuclear mean-field models. Exotic nuclei, and particularly SHE, may
help towards an answer.

Concerning the incompressibility $K_\infty$, the SHF forces
almost all gather nicely around the generally accepted value of
$230\,{\rm MeV}$ \cite{incomp}, SkP being an exception with a
rather low value of $K_\infty$. The RMF forces make quite different
predictions. NL-Z produces too low $K_\infty$, which results from the
fit, while NL3 comes up with a rather large value, which is to some extent a
bias entering the adjustment. The actual number is probably at the
upper edge of presently accepted values. A similarly large value is
produced by TM1.

The largest variations are seen for the asymmetry energy $a_{\rm
sym}$. The LDM predicts values around $30\,{\rm MeV}$. Indeed, most
SHF forces reproduce that nicely, with the exception of SkI3 which
comes out too high and Z$_\sigma$ which yields a somewhat low value,  but
the RMF forces generally yield a very large value for $a_{\rm
sym}$. One then wonders what the properties of the isovector dipole
giant resonance might be. It turns out that its position depends not
only on $a_{\rm sym}$ but also on the isovector effective mass, or sum
rule enhancement factor $\kappa$, respectively. Most Skyrme forces have
rather low $\kappa\approx 0\!-\!0.25$, yielding correct resonance
frequencies for $a_{\rm sym}\approx 30\,{\rm MeV}$. The RMF forces
have much larger $\kappa\approx 0.75$ and here the value $a_{\rm
sym}\approx 40\,{\rm MeV}$ is appropriate. For a detailed discussion
of these somewhat surprising interconnections see \cite{varenna}.  It
remains that there is a substantial difference between SHF and RMF in
that respect. The reasons are as yet unclear; it is probably again caused
by the different form of density-dependence. 
%
%
\subsection{Binding Energies of Superheavy Nuclei}
%
%
\begin{figure}[t!]
\centerline{\epsfig{figure=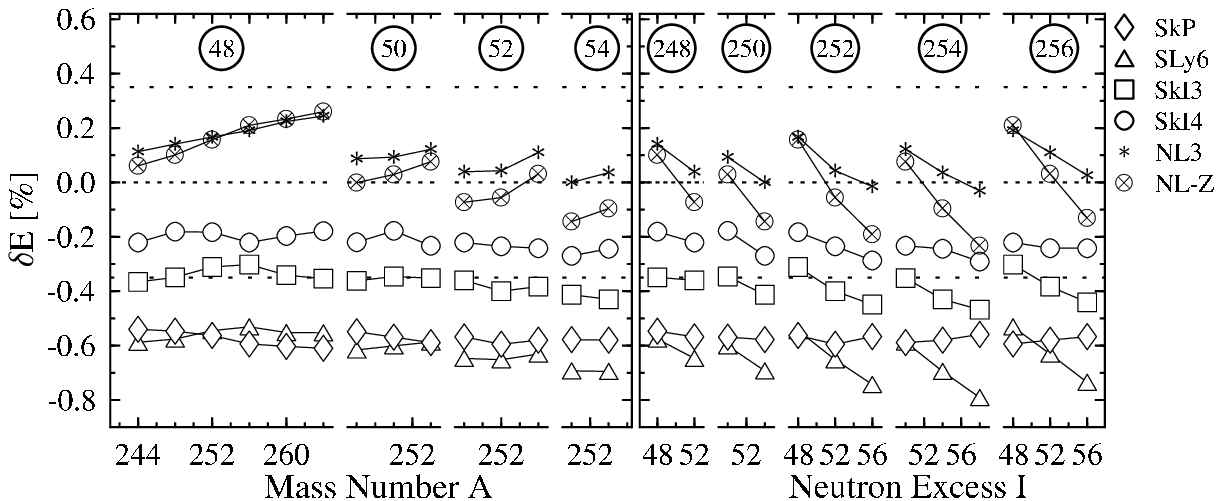,width=15.0cm}}
\caption{\label{fig:deltae_a}
Relative error on binding energies 
\mbox{$\delta E = (E_{\rm calc}-E_{\rm expt})/E_{\rm expt}$},
of even-even superheavy nuclei  
for the selection of forces drawn versus mass number 
\mbox{$A = N + Z$} for constant \mbox{$I = N - Z$} (left panel) and
versus $I$ for constant $A$ (right panel). Note that the left panel
corresponds to $\alpha$--decay chains. The quality margin of $0.35 \; \%$
which was the achieved average error for normal nuclei is indicated
by dotted lines. See Ref.~\protect\cite{Bue98a} for a discussion 
of the theoretical uncertainties. Data taken from 
\protect\cite{Bue98a,dubna}.
}
\end{figure}
%
%
We now proceed to the discussion of SHE. The first feature to look at
is, of course, the binding energy. Fig.~\ref{fig:deltae_a} 
shows the relative error on binding energies $\delta E$
for a selection of already known SHE. 
One sees at first glance, that the errors stretch out
towards under-binding. The RMF forces remain very well within the
desired error bands. The two SHF forces with extended spin-orbit
splitting also stay just within the bounds, and all
conventional SHF forces fall below the $0.35 \; \%$ margin. This is most
probably not caused by the underlying bulk properties but related to
shell effects. Note that Fig.~\ref{fig:deltae_a} presents the same 
data in two different fashions to disentangle different trends
in the error stemming from the isoscalar (\mbox{$I =$} const.) and 
isovector (\mbox{$A=$} const) channel of the interaction \cite{dubna}. 
We look first at the left panel where the trends with $A$ are drawn. It is 
gratifying to see that all SHF forces basically follow a horizontal 
line which implies that the isoscalar bulk properties are described
correctly. The RMF lines, however, have visible slopes, showing
that the trends with $A$ are not perfectly reproduced. Such a feature
had already been hinted at in Fig.~\ref{fig:quality} where the volume
parameters $a_{\rm vol}$ and $\rho_0$ from the RMF differed from those
of SHF and from the typical LDM values. This again most probably
indicates a deficiency of the density-dependence in RMF.

The right panel of Fig.~\ref{fig:deltae_a} displays the isovector
trends. Clearly almost no force hits these trends
correctly. SkP shows the most horizontal lines and thus seems to 
incorporate some correct isovector features.
It is, on the other hand, a strange surprise that
SLy6 deviates so much from the experimental isovector trends. This
force was intended to perform particularly well in the isovector
channel. The feature has yet to be fully understood. Keeping in mind that the
actual trends are a mix of isovector bulk properties and shell
effects it is most probable that the shell effects cause these deviations.
SkI3 has as bad trends as SLy6 while SkI4 performs a bit better. This
again is an accident because these trends had not been included in the
fit.  The RMF forces also fail with respect to isovector trends,
The force NL3 performing a bit better than NL-Z, possibly because
isovector trends of binding energies had been included in the
fit. It is even more surprising that NL3 does not perform better.
There are still open problems with a proper parametrisation of the
isovector channel in the RMF. Remembering that there is only one isovector
field taken into account, one would like to also
incorporate the scalar-isovector field (the $\delta$ meson) to achieve a
better isovector performance in the RMF, but this channel 
probably needs non-linear couplings because a simple linear ansatz
did not lead to improvements \cite{RutzDiss}.

All these results on this apparently innocuous observable binding energy hint
that new information from SHE sheds new light on mean-field
models. A thorough study of the reasons for underbinding and 
unresolved trends has yet to come and will certainly help to 
deduce new constraints on the parametrisations.

%
%
\subsection{Shell Effects}
%
%
\begin{figure}[t!]
\centerline{\epsfig{figure=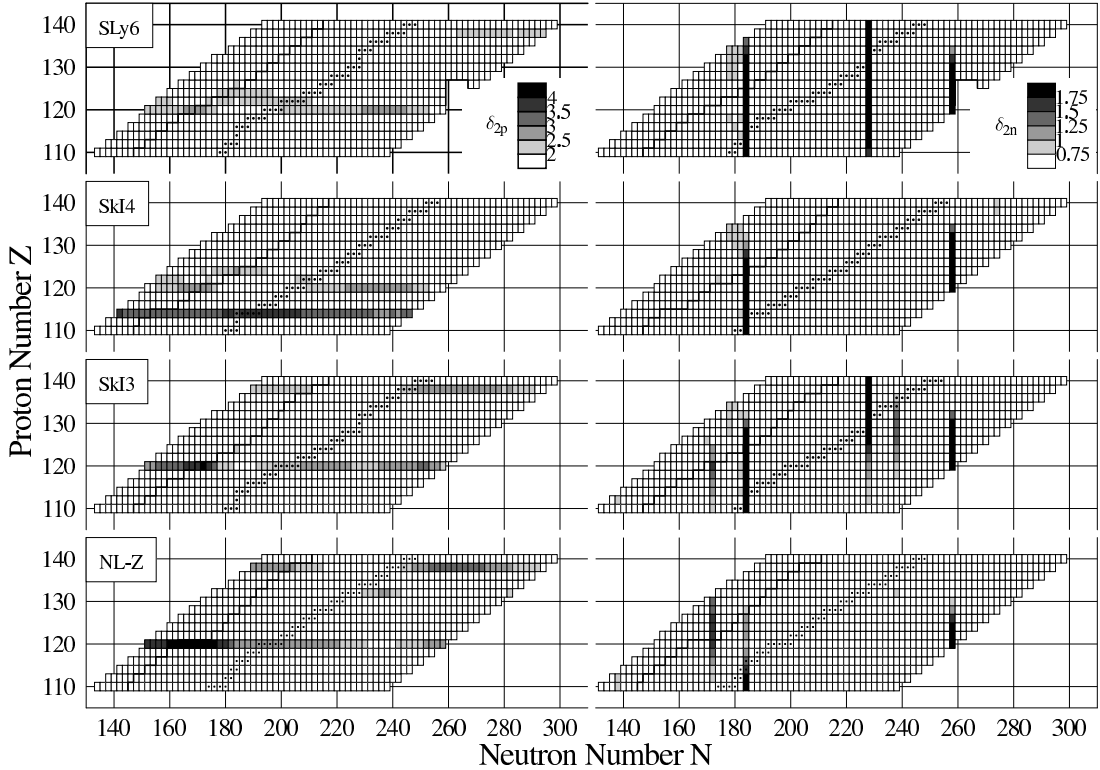,width=14.5cm}}
\caption{\label{fig:spheres}
Grey scale plots of two-proton shell gaps $\delta_{\rm 2p}$ (left column) 
and two-neutron shell gaps $\delta_{\rm 2n}$ (right column) in the $N$--$Z$ 
plane for spherical configurations
calculated with the effective interactions as indicated. The 
assignment of scales differs for protons and neutrons, see the 
uppermost boxes where the scales are indicated in units of MeV.
The most bound nuclei in each isobaric chain and the two-proton
drip-line are emphasized. Data taken from \protect\cite{firstSH}.
}
\end{figure}
%
%
Shell effects are constitutive for the existence of SHE and they
play a crucial role in determining the actual stability against fission. 
It is thus worthwhile to have a closer look at shell effects. A prominent 
feature is the occurrence of shell closures or magic numbers, respectively, 
in the single-particle spectrum. One way to characterise
them is to examine the two-nucleon separation energies, e.g.\ 
\mbox{$S_{2n}=E(Z,N-2)-E(Z,N)$}. They display a sudden drop at shell
closures because it is easier to remove nucleons from the next open
shell (the former valence shell). The size of the step is a measure
for the ``magicity'' of the shell closure. It is given by the
two-nucleon shell gaps, e.g.\ for the neutrons
\begin{eqnarray}
\delta_{\rm 2n} (Z,N)
& = & S_{\rm 2n} (Z,N) - S_{\rm 2n} (Z,N+2)
      \nonumber \\
& = & E (Z,N+2) - 2 E (Z,N) + E (Z,N-2)
\end{eqnarray} 
and similarly for the protons. This quantity is a way to access the
gap between last occupied and first unoccupied single-particle states,
see e.g.\ \cite{Ben99a}.
Peaks in $\delta_{2p}$ or $\delta_{2n}$ indicate a shell closure. 
Fig.~\ref{fig:spheres} shows proton and neutron shell gaps for a large
range of SHE and for a subselection of forces. It is done for 
simplicity with spherical calculations. This suffices when searching
for spherical shell closures. Deformation might change 
the picture in details and adds deformed shell closures,
e.g.\ \mbox{$N=162$} or \mbox{$Z=108$}, see \cite{Bue98a}.
The left panels show $\delta_{2p}$. The dark horizontal stripes thus
indicate the closed proton shells.  The right panels show
$\delta_{2n}$, the dark vertical stripes there stay for closed
neutron shells. The different forces show quite different patterns.
This holds particularly for the proton shell closures. The RMF force
NL--Z and the most RMF-like SHF force SkI3 predict a magic \mbox{$Z = 120$}
whereas SkI4 prefers \mbox{$Z = 114$} and SkP shows no pronounced proton
shell closures at all. For the neutrons, all SHF forces predict a 
\mbox{$N=184$} shell while RMF prefers \mbox{$N=172$}. That is not 
mutually exclusive. Several forces, SHF and RMF, have both closures.
%
%
\begin{figure}[t!]
\centerline{\epsfig{figure=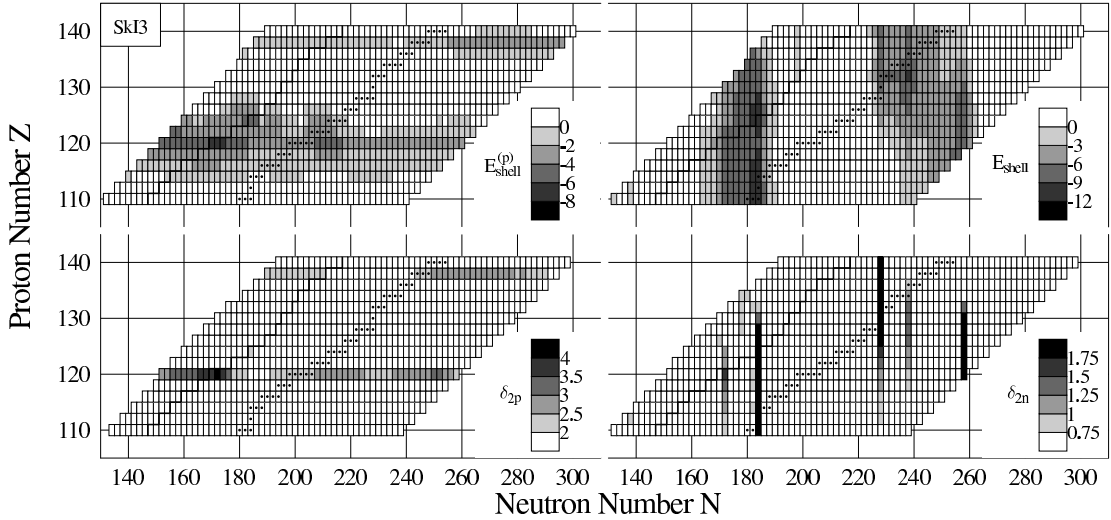}}
\caption{\label{fig:eshell}
Grey-scale plots of the shell correction energies $E_{\rm shell}^{(q)}$
(upper panels) and the two-nucleon shell gaps $\delta_{2q}$ (lower panels)
for protons (left panels) and neutrons (right panels) calculated
with SkI3 for spherical shapes. 
Data taken from \protect\cite{firstSH,newshelcor}.
}
\end{figure}
%
%
%
%
\begin{figure}[b!]
\centerline{\epsfig{figure=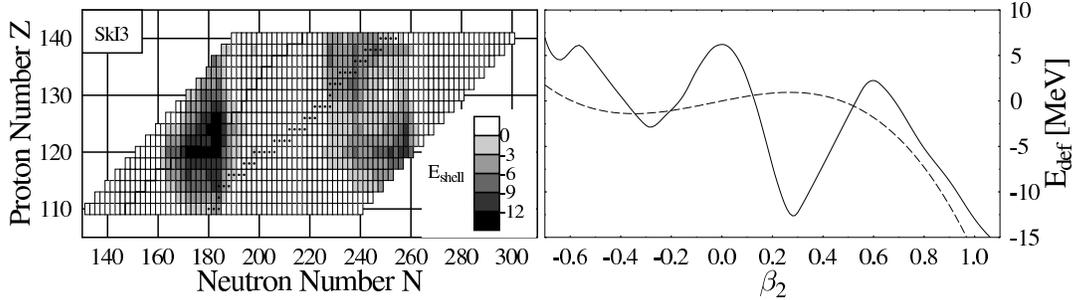}}
\caption{\label{fig:eshell2}
Grey-scale plots of the total shell correction energy 
\mbox{$E_{\rm shell} = E_{\rm shell}^{\rm (p)} + E_{\rm shell}^{\rm (n)}$}
calculated with SkI3 for spherical shapes (left panel) and 
schematic plot of the shell stabilization of a superheavy nucleus.
(right panel).
Data taken from \protect\cite{newshelcor}.
}
\end{figure}
%
%

The shell gaps $\delta_{2q}$ are very useful when searching shell
closures, but they are not directly related to the ``shell effect''
that stabilizes SHE against Coulomb fission. This quantity is provided 
by the shell correction energy
\begin{equation}
E_{\rm shell} 
= \sum_{\alpha} \varepsilon_\alpha 
  - \int \! {\rm d} \varepsilon \; \tilde{g} (\varepsilon)
\quad.
\end{equation}
High level density around the Fermi surface yields positive $E_{\rm shell}$ 
which corresponds to reduced binding. Smaller--than--average level 
density, in turn, corresponds to negative $E_{\rm shell}$, i.e.\ extra 
binding from shell effects \cite{LDM,micmac,micmac2}. 
Fig.~\ref{fig:eshell} shows an example of the individual shell correction
of protons and neutrons in comparison with the two-nucleon shell
gaps $\delta_{2q}$. While in mac--mic models the shell correction 
is an constitutive part of the calculation of the binding energy, the
values presented here are a posteriori analysed 
from the actual single-particle 
spectra of fully self-consistent calculations as a measure of 
the shell effect \cite{shelcor}.  
Maximum (negative) values of the shell corrections 
coincide with the peaks in $\delta_{2q}$, 
but there is also a significant difference. While the two-nucleon 
shell gaps show isolated peaks, the shell corrections appear as 
rather broad valleys of shell stabilised nuclei. 
The valley is broader than that around magic shells for normal nuclei.
The stabilizing effect of the shell correction is given by the sum of
the shell corrections for proton and neutrons. The mechanism is
sketched in the right panel of Fig.~\ref{fig:eshell2}. The dashed line
indicates the smooth deformation energy curve corresponding to the LDM
background. It is repulsive for SHE which means that they all would be
fission-unstable in a LDM world. The full line has the shell
corrections added. They oscillate with deformation and this generates
minima which are stabilised against fission. The amplitude of the
oscillations corresponds to the height of the fission barrier. Thus
the depth of the shell-correction valley is a rough measure for
fission stability. The total shell correction energy is given in the
left panel of Fig.~\ref{fig:eshell2}. There emerges a broad region of
shell stabilised SHE. The positive aspect of these findings is that
one has good chances to hit long-living SHE in a variety of entrance
channels. The negative aspect is that the quest for doubly-magic SHE is
misleading. Magicity is not very pronounced out there, a feature which
was already seen in mac-mic models \cite{FY}. The crucial
features for SHE are large shell corrections, and these exist; even
better, they appear for a broad range of $Z$ and $N$ which makes the
search for SHE in some sense comfortable.
%
%
\subsection{Single--particle structure}
\begin{figure}[t!]
\centerline{\epsfig{figure=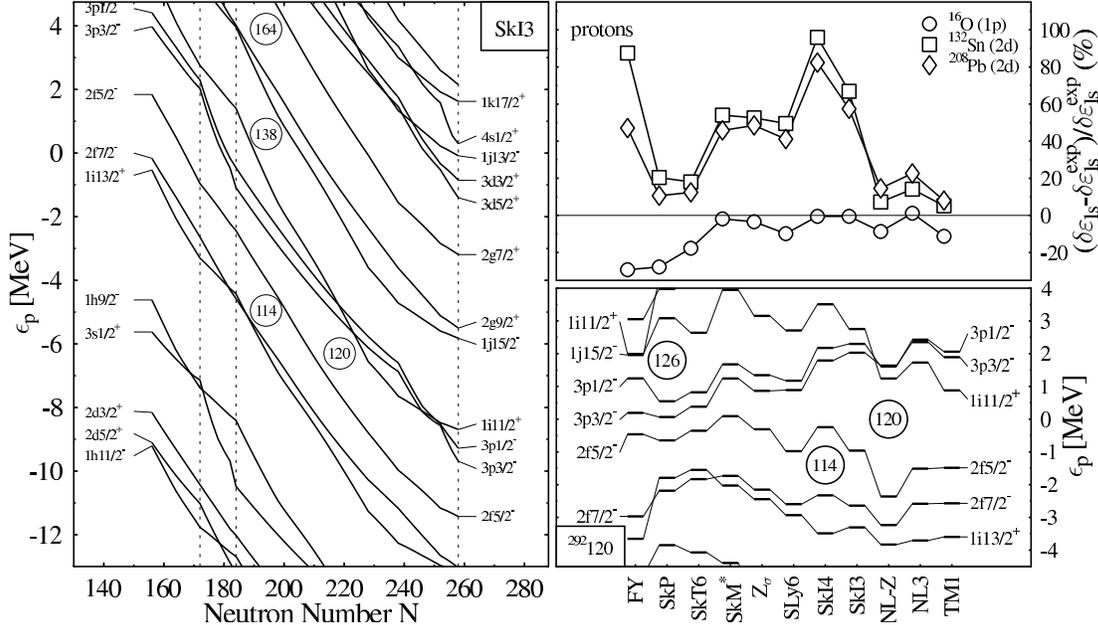,width=14.5cm}}
\caption{\label{fig:shells}
Left panel:
Single-proton levels in the vicinity of the Fermi
energy for the chain of \mbox{$Z = 120$} isotopes as predicted 
by SkI3 plotted versus the neutron number. The dotted 
vertical lines indicate the spherical magic neutron numbers
\mbox{$N=172$}, \mbox{$N=184$} and \mbox{$N=258$} predicted
by SkI3.
Right top panel: 
Relative error of the spin-orbit splitting of selected
proton states in the doubly-magic nuclei as indicated.
Experimental data taken from \protect\cite{NUDAT} except
the proton $2d$ splitting in $^{132}$Sn taken from 
\protect\cite{Sb133}. Note that earlier papers 
\protect\cite{Ben99a} have used the larger value of this 
splitting taken from \protect\cite{NUDAT}. FY denotes
the Folded--Yukawa single-particle potential widely
used in mac--mic models \protect\cite{FRDM}.
Right bottom panel:
Single-nucleon spectra for $^{292}120$ at spherical shape 
calculated with the interactions as indicated
Left: proton spectra. Right: neutron spectra.
Note that this nucleus is deformed for most SHF interactions.
Data taken from \protect\cite{Ben99a}.
}
\end{figure}

Both previous sections have pointed out signatures of shell effects. 
In this section we sketch the actual single-particle spectra of SHE.
The left panel of Fig.~\ref{fig:shells} shows the proton levels near 
the Fermi surface for 
\mbox{$Z=120$} and varying $N$. The overall trend is obvious: proton 
levels become more deeply bound with increasing $N$, but note the
change of the gap at \mbox{$Z=120$} along the isotopic chain which is
coupled to the magic neutron number; a slight shift of the 
single-particle energies destroys the \mbox{$Z=120$} around 
\mbox{$N=184$}. This illustrates the same effect seen in 
the $\delta_{\rm 2p}$ in Fig.~\ref{fig:spheres}.
It is a typical self-consistency
effect not seen in earlier mac-mic calculations caused by the strong 
coupling of bulk properties and single-particle structure. We will
come back to that in Section~\ref{Subsect:DensityProfiles}.
The right lower panel of Fig.~\ref{fig:shells} tries 
to visualize the competition between the \mbox{$Z=114$} and 
\mbox{$Z=120$} shell.
%
%
%
%

The lower right panel of \ref{fig:shells} displays the proton spectra 
for $^{192}120$. Most interactions predict the same level-ordering
in the superheavy region, the different bias on 114 and 120 among the
forces found in Fig~\ref{fig:spheres} is related to slight changes 
in the relative distances of the levels. 
The reason for this behaviour is clearly apparent: the \mbox{$Z=114$} 
shell corresponds to large spin-orbit splitting of the $2f$ proton levels, 
while the \mbox{$Z=120$} shell requires small $2f$ splitting. 
Because the self-consistency makes the level scheme depend strongly 
on $Z$ and $N$ (cf. the left panel of Fig.~\ref{fig:shells})
this graph is by itself not fully conclusive, but a more careful 
examination shows the conclusion to be valid \cite{Ben99a}. 

It is thus the spin-orbit force which decides on the preferred shell
closure. To estimate the reliability of the spin-orbit splitting, we
look at its performance in normal nuclei, see the right upper panel in
Fig.~\ref{fig:shells}. It shows the relative error in spin-orbit
splittings in selected proton levels of doubly-magic nuclei (only
``safe'' splittings have been chosen according to the study of
\cite{Rut98a}). There is a systematic difference between SHF and RMF.
The RMF forces give a very satisfactory description of the data
in all cases which emerges without any special fit to spectral data. 
It is a natural outcome of the Dirac equation combined with two 
strong fields (scalar versus vector) which counteract in the potential 
but cooperate in the spin-orbit force \cite{duerr}.
The non-relativistic interactions fall into two groups which can
be distinguished by their performance for spin-orbit splittings:
those where the spin-orbit interaction is adjusted to several 
nuclei throughout the chart of nuclei (FY, SkP, SkT6) and those
fitted solely to $^{16}$O. 
The latter reproduce the $1p$ proton splitting in $^{16}$O
but overestimate the splittings in heavier nuclei. This is an
unpleasant common feature of all non-relativistic models. Owing
to the fit strategy  the errors are centered around zero for 
SkP and SkT6 which gives a better overall performance but does
not cure the problem.
The other forces show even more serious discrepancies, 
particularly SkI4. This makes the (among self-consistent models)
unique prediction of a 
spherical \mbox{$Z=114$} shell (that is directly related to 
the spin-orbit splitting) of this interaction very questionable.
This is not necessarily a defect of the SHF as such. 
Better performing parametrizations are feasible
but have yet to be fully worked out. The mismatch in the spin-orbit
splittings should be a warning that extrapolations to 
detailed features of SHE have to be
taken with care because these depend sensitively on shell effects.
Note in that context that the much celebrated mic-mac
approach gives also questionable predictions for shell closures 
as it displays also rather large errors, see the column FY 
in Fig.~\ref{fig:shells}. 
%
%
\subsection{Density Profiles}
\label{Subsect:DensityProfiles}
%
%
\begin{figure}[t!]
\centerline{\epsfig{figure=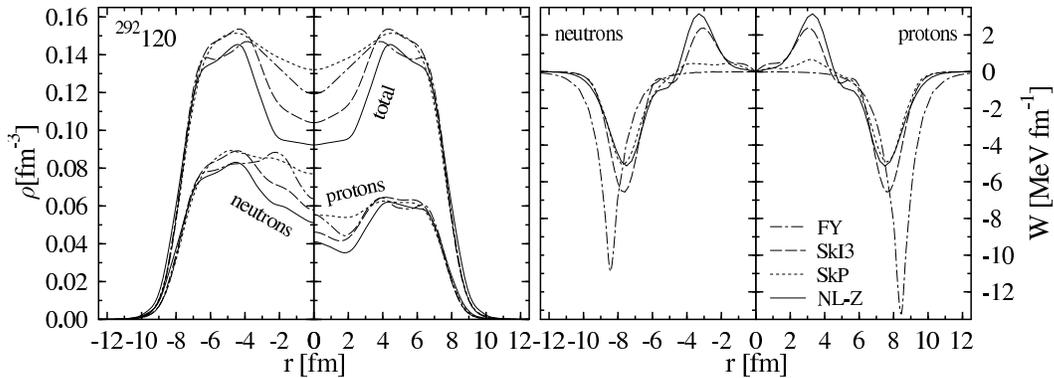}}
\caption{\label{fig:dens}
Density distribution and radial component of the spin-orbit potential 
$W_r$ of protons and neutrons in ${}^{292}_{172}120$, calculated with 
the forces a indicated for spherical configurations.
The total density is plotted in the left panels as well.The density 
distributions calculated from the single-particle wave functions 
as they come out in the FY model are drawn for comparison. All models
except SkP show a central depression in the density distribution, This
has a visible impact on the spin-orbit potential which is proportional
to the gradient of the densities.
Taken from \protect\cite{Ben99a}.
}
\end{figure}
%
%
The unusual spin-orbit splitting for $^{292}120$ seen in the lower 
right panel of Fig.~\ref{fig:shells} is related to an unusual density 
profile of this nucleus, see the left panel of Fig.~\ref{fig:dens}.
The pattern can be understood as a cooperative effect from Coulomb repulsion 
and shell fluctuations (where the shell effect actually takes the lead).
The dip at the center may be understood at first glance from Coulomb
repulsion. But note that the depth of the dip differs substantially
amongst the forces. Mean-field interactions with effective mass 
$m^*/m=1$ (SkP) show only a shallow minimum whereas those with low $m^*/m$
display a deep central depression, half-way to a bubble nucleus 
\cite{bubble}. This is perfectly consistent with a shell fluctuation 
giving rise to the oscillation of the density in the interior 
\cite{shellflu} and the low effective masses 
in the RMF make particularly large fluctuations.
Shell structure thus dominates the Coulomb effects on the densities
which is confirmed looking at the FY predictions where this effects appears
even without the self-consistent feedback between densities and
potentials. 

The right panels of Fig.~\ref{fig:dens} show the spin-orbit potentials
$W_r$ for $^{292}120$. The dominant part of $W_r$ is located at the surface, 
\mbox{$r \approx 8 \, {\rm fm}$} where we see that SkI3 has a larger 
amplitude. That agrees with the larger spin-orbit splittings
found in the upper right panel of Fig.~\ref{fig:shells}. At second
glance, we see that the maximum of $|W|$ is shifted to smaller $r$ for
the RMF force NL--Z. This is a tiny, but systematic, effect in all
spin-orbit potentials which we have looked at, and is probably one
ingredient for the superior performance of the RMF with respect to
spin-orbit splittings. As $\vec{W}$ is approximately $\propto \nabla\rho$
in self-consistent models, it directly reflects the shell oscillation 
of the density distribution. The central 
depression of the density leads to the large positive peak in $W_r$
around \mbox{$r \approx 3 \, {\rm fm}$}. This may lead to a
disappearance of the splitting for states with low $l$ (which are sensitive 
to the interior) that is crucial for the appearance of the spherical 
\mbox{$Z=120$} and \mbox{$N=172$} shells in some of the models
(sometimes even inversion of sign, see the proton $3p$ 
states in the lower right panel of Fig.~\ref{fig:shells} as predicted
by NL--Z, NL3 and TM1). Again this is a self-consistent effect which 
cannot be described with (current) mic-mac models. 
The spin-orbit potential
$W_r$ from the FY model is proportional to the gradient 
of the parameterized average potential and that gradient disappears 
inside the nucleus. 
Note that in this case the peak of $W_r$ is at larger radii, much
narrower and of larger amplitude than in all self-consistent models
which causes the differences in the spin-orbit splittings visible
in Fig.~\ref{fig:shells}. 
%
%
\subsection{Potential Energy Surfaces for Fission}
%
%
\begin{figure}[t!]
\centerline{\epsfig{figure=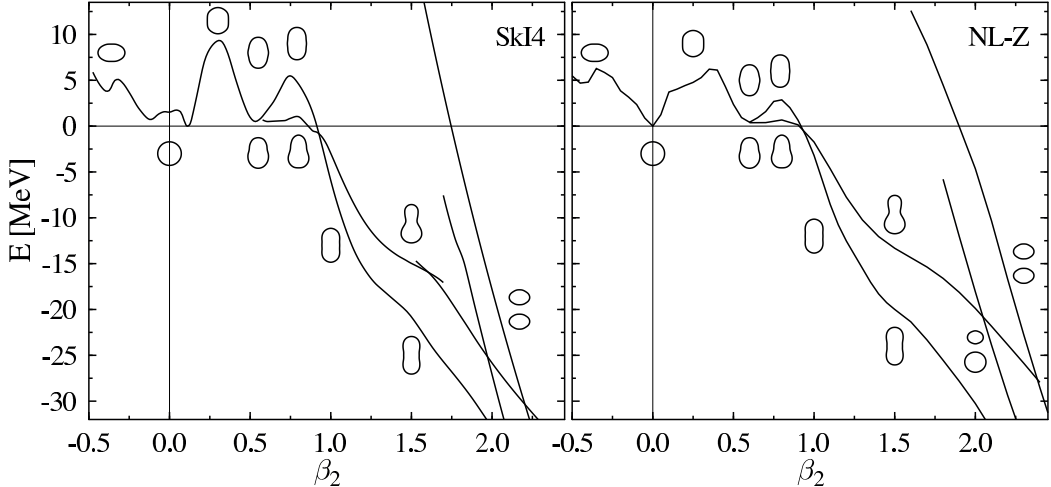}}
\caption{\label{fig:fiss}
Valleys in the PES of the potential doubly magic nucleus
${}^{292}_{172}{120}$, calculated with SkI4 (left panel) and NL--Z 
(right panel) in axial symmetry. Results from calculations in different 
symmetries can be distinguished by the mass density contours which are
drawn near the corresponding curves.
Data taken from \protect\cite{PESfiss}.
}
\end{figure}
%
%
Fig.~\ref{fig:fiss} shows the deformation energy curves, usually
called potential-energy surfaces (PES), for fission of $^{292}120$.
For large deformations \mbox{$|\beta_2| > 0.5$} both
models give very similar predictions, obvious differences (related to
spherical shell structure) appear for smaller deformations only. Small
deformations (\mbox{$\beta_2 < 0.5$}) unambiguously prefer
(reflection-) symmetric shapes. Distinct symmetric and asymmetric
fission paths develop for larger $\beta_2$, which is a general feature
of very heavy nuclei \cite{PESfiss}. Both also show a very low second
minimum, in fact rather a saddle point, since there is no second
barrier. A very interesting feature is the octupole softness around
\mbox{$\beta_2 \approx 0.5$}, the PES in octupole direction is almost
perfectly flat between \mbox{$\beta_3 = 0.0$} and \mbox{$\beta_3
\approx 0.5$}.  The asymmetric path continues then with negligible barrier.  
This is very different from the pronounced double-humped structure of 
actinide nuclei. For some lighter nuclides the asymmetric path even 
lowers the first barrier as well as triaxial shapes do
\protect\cite{PESfiss,Cwi96a}.  The inner barrier is still very high,
showing that SHE could very well be relatively stable against
spontaneous fission.  All of these are generic features of fission
paths in shell-stabilized SHE \cite{PESfiss}. There are, of course,
some differences between the two forces shown in
Fig.~\ref{fig:fiss}. \mbox{$Z=120$} is a small closure for NL--Z and
accordingly we have a well developed spherical minimum, but SkI4
already produces a deformed minimum. Note that this does little harm
to the stability. The first barrier is even higher than for NL--Z. As
seen already in Fig.~\ref{fig:eshell}, shell stabilisation works very
well even somewhat remote from spherical shell closures and even for
deformed shapes \cite{Ben00a}.

The PES of SkI4 is instructive in another respect.
We see at least three almost degenerate minima. That is a typical example of
shape coexistence. And it demonstrates once more that exotic nuclei are very
likely to display shape coexistence \cite{shapeiso,186Pb}.
The example of $^{292}120$ with NL--Z showing a clear cut spherical minimum
is rather the exception than the rule, see also \cite{Ber96a}.
%
%
\subsection{Recent $\alpha$--Decay Chains}
\begin{figure}[t!]
\centerline{
\epsfig{file=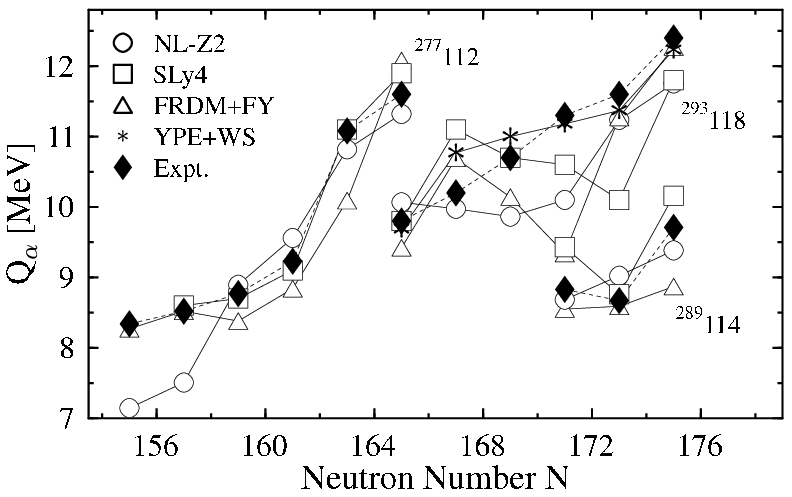}
\epsfig{file=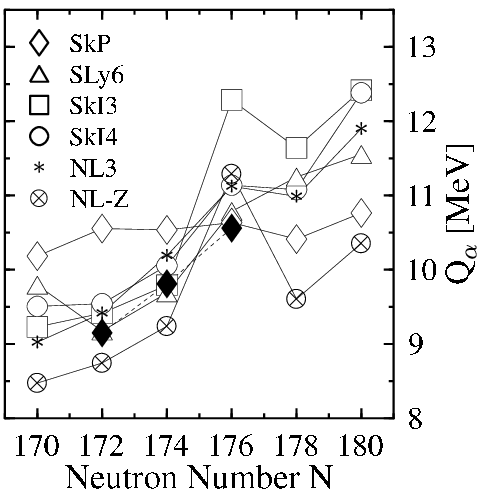}
}
\caption{\label{Fig:ug:Q:alpha}
Comparison of experimental and calculated $Q_\alpha$ values
for the decay chains of $^{277}_{165}112$, $^{289}_{175}114$,
and $^{293}_{175}118$ (left panel) and the $\alpha$--decay chain 
leading through $^{292}_{176}$116 (right panel), in the latter 
three cases following the mass and charge assignment of the 
experimental groups. For odd-$N$ nuclei the calculated values 
from SLy4 and NL--Z2 connect the lowest states with positive parity 
in all cases, while the FRDM+FY and YPE+WS data are ground state 
to ground state values. 
Data taken from \protect\cite{Ben00a,dubna}.
}
\end{figure}

The preferred decay mode of shell stabilised SHE is $\alpha$ decay.
A key quantity there is the $Q_\alpha$ value for the reaction which
is defined as 
\begin{equation}
Q_\alpha (N,Z)
= E(N,Z) - E(N-2,Z-2) - E (2,2)
\end{equation}
Recent experiments \cite{Z114,Z116,Z118} have reached the lower bounds 
of the island of spherical SHE which is expected somewhere around
$114 \leq Z \leq 126$ and $172 \leq N \leq 184$ depending on the model,
as discussed above.
It is interesting to compare the new data on $Q_\alpha$ with 
predictions from mean-field models. 
As $\alpha$--decay chains have constant \mbox{$I = N-Z$}, 
the isoscalar channel mainly determines the slope
of the $Q_\alpha$ and the isovector channel the offset.
Shell effects bend the curves locally, leading to kinks and peaks. 
Recent investigations of $Q_\alpha$ throughout the region of SHE 
with SLy4 in \cite{Cwi99a} and NL--Z2 in \cite{Ben00a} 
show a good overall description of the data by these two
forces, although none of the interactions reproduces all 
details of the data. Most of the recently synthesized
SHE are odd-$A$ nuclei where the unpaired nucleon complicates the
theoretical description, see \cite{Cwi99a,Ben00a,Cwi94a}.
Fig.~\ref{Fig:ug:Q:alpha} compares data with calculations in 
the self-consistent SHF and RMF models (using the forces SLy4 
and NL--Z2 respectively) and the mac-mic FRDM+FY and YPE+WS models.
In view of the uncertainties, SLy4 and NL--Z2 give a very good
description of the data for the decay chain of $^{277}_{165}112$ 
and reproduce the \mbox{$N=162$} shell effect,
which cannot be seen in the FRDM+FY predictions. While all models
give similar predictions for this well-established chain, the spread
among the models is much larger for the new chains. All models with the 
exception of macroscopic YPE+WS model show spherical or deformed shells 
which cannot be seen in the data.
The right panel of Fig.~\ref{Fig:ug:Q:alpha}
compares predictions with the recent data for the even-even 
$^{292}_{176}$116 decay chain (which still have to be viewed as 
preliminary). It is most interesting that the data agree with calculated 
values from interactions SkI4, SLy6 and NL3, although these
three forces make different predictions for the spherical magic
numbers, i.e.\ SkI4 (\mbox{$Z=114$}, \mbox{$N=184$}), SLy6 
(\mbox{$Z=126$}, \mbox{$N=184$}), and NL3 (\mbox{$Z=120$}, \mbox{$N=172$}). 
All other interactions show wrong overall trends of 
the $Q_\alpha$ or pronounced deformed shells in disagreement with the 
data or even both. This demonstrates that predictions for spherical shell 
closures and binding energy systematics are fairly independent. 
%
%
%
%

The experimental values follow a smooth trend while most mean-field results 
have a pronounced kink at \mbox{$N=176$}. This is a shell effect related
to predicted deformed shells which is not reflected in the data. The
resolution of that puzzle lies in correlation effects. The PES of
these SHE are rather soft (see also the previous subsection). The
ground state then explores large fluctuations in quadrupole
deformation $\beta_2$ and the pure mean-field minimum is insufficient in
such a situation. One has at least to evaluate the correlation effects
for $\beta_2$ motion, for an example from stable nuclei see e.g.\
\cite{girod}. Applying such a scheme to the chain of SHE indeed yields
a smooth trend of the $Q_\alpha$ \cite{Rei00a}, but this aspect of
correlations goes beyond the scope of this paper.
%
%
\section{Conclusions}
We have reviewed self-consistent models for nuclear structure, thereby
concentrating on the two most widely used brands: the relativistic
mean-field model (RMF) and the non-relativistic Skyrme--Hartree--Fock
model (SHF). A brief discussion of the formal properties has shown the
relation between these two models and to nuclear bulk properties.
Each term in the SHF energy can be directly connected with a bulk
property (as e.g. volume energy, asymmetry energy, etc). An exception
is, of course, the spin-orbit term which disappears in bulk matter.
The RMF can be connected with SHF by virtue of a non-relativistic
expansion in orders of $v/c$ and in orders of meson range. The basic
structures are fully comparable whereas the details of density
dependence differ. An advantage of the RMF is that the spin-orbit
force is automatically included without the need of separate tuning. 
The SHF, on the other hand, is superior in its flexibility to
accommodate isovector forces.
Both models contain a good handful of free parameters which need
to be adjusted phenomenologically. Different demands and bias in the
adjustment has led to a world of different forces, in SHF as well as
in RMF. We have selected an overseeable set of typical forces to
display the possible variations in the results. 
The basic properties for normal nuclei are all very well reproduced
by all chosen forces. An interesting detail is the isotope shift of
radii in Pb. The remarked kink at the magic $^{208}$Pb is immediately
reproduced by the RMF but can only be described within SHF after an
appropriate extension of the (isovector) spin-orbit force.

The discussion of results has concentrated on superheavy elements
(SHE). The average error of binding energies in known SHE covers a
larger span than the error in stable nuclei. This is an expected
result because extrapolations always tend to scatter the errors.
The positive aspect is that all errors remain in bearable bounds
(safely below 1 $\%$) and that there remain even several forces 
which maintain the quality found in stable nuclei. 
A different feature is given by looking at the relative errors of
binding energies nucleus by nucleus, thus displaying the trends in
these errors.  The quality in reproducing the trends is found to be
independent from the average quality of the binding energies. Large
differences between the forces are seen for the trends in mass number
$A$ and neutron excess $N-Z$ where the SHF forces generally perform
better with respect to these trends.
A different way to look at trends is provided by the two-nucleon
separation energies and $Q_\alpha$ values.
The  $Q_\alpha$ for recently discovered chains of SHE are nicely
reproduced within $\pm 1$ MeV for the more recent and well
adjusted parametrisations in SHF and RMF. Somewhat more variance 
is found for the separation energies (not shown in this paper).
The trend of the trends is given with the two-nucleon shell gaps,
i.e.\ the difference of adjacent separation energies. They depend
predominantly on shell structure (unlike binding and separation
energies which are also influenced by the bulk properties of a force).
Large gaps serve as an indicator for magic shell closures.
The predictions on the two-nucleon shell gaps vary
substantially, to the extend that different forces predict shell
closures at different proton numbers. The differences look less
dramatic if one realises that the overall size of the two-nucleon shell
gaps is small in any case. The concept of magic shell closures seems
to fade away in SHE.
One has to remind that magic shells had been looked for as a simple
guideline where to find SHE which are sufficiently shell-stabilised
against Coulomb pressure towards spontaneous fission. The ultimate, but
hard to evaluate, criterion is the height of the fission
barrier. Simpler to compute is the shell-correction energy which can
serve as a rough estimate: large negative shell-correction energy is a
necessary condition for stability against fission.  We find for all
forces a broad valley of large shell correction energies. This a
welcome feature as it leaves some freedom in the choice of the
reaction channels. Thus we see good chances to hit many more SHE in
near future experiments, in spite of the fact that pronounced doubly
magic systems will not be found.

The investigations have demonstrated the high descriptive power of
nowadays mean-field models. They have also revealed some weak points
where further fine-tuning is needed, taking advantage of the many new
data from exotic nuclei in general and SHE in particular. Last not
least, one explores the limits of mean-field models when going towards
the limits of stability. Shape coexistence and subsequent need for
correlation effects shows up notoriously for the less well bound
nuclei.
%
%
\subsection*{Acknowledgements}
We would like to thank our collaborators 
T.\ B\"urvenich, 
S.\ {\'C}wiok, 
D.~J.\ Dean, 
J.\ Dobaczewski,
P.\ Fleischer, 
W.\ Greiner, 
A.\ Kruppa, 
W.\ Nazarewicz, 
Ch.\ Rei{\ss},
K.\ Rutz,
T.\ Schilling,
M.~R.\ Strayer,
and
T.\ Vertse
for their contributions and helpful hints.
We also acknowledge many inspiring discussions with our
experimental colleagues
J.\ Friedrich,
S.\ Hofmann
G.\ M\"unzenberg,
V.\ Ninov,
and 
Yu.\ Ts.\ Oganessian.
This work was supported by Bundesministerium f\"ur
Bildung und Forschung (BMBF), Project No.\ 06 ER 808 and by
Gesellschaft f\"ur Schwerionenforschung (GSI).
The Joint Institute for Heavy Ion Research has as member institutions 
the University of Tennessee, Vanderbilt University, and the Oak Ridge 
National Laboratory; it is supported by the members and by the Department 
of Energy through Contract No.\ DE--FG05--87ER40361 with the University of 
Tennessee. 
%
%


\begin{thebibliography}{999}

\bibitem{LDM}
  S. G. Nilsson, I. Ragnarsson,
  \emph{Shapes and Shells in Nuclear Structure},
  Cambridge University Press, 1995.

\bibitem{micmac}
  M. Brack, J. Damg{\aa}rd, A. S. Jensen, H. C. Pauli, V. M. Strutinsky,
  C. Y. Wong, 
  Rev. Mod. Phys. \textbf{44}, 320 (1972).

\bibitem{micmac2}
  M. Brack, 
  Proc. of the ``International Workshop on Nuclear Structure Models'',
  Oak Ridge, Tennessee, March 16--25, 1992,
  edited by R. Bengtsson, J. Draayer, W. Nazarewicz,
  World Scientific, Singapore, 1992, page 165.

\bibitem{Mosel} 
  U. Mosel, W. Greiner, 
  Z. Phys. {\bf 222}, 261 (1969).

\bibitem{SuperNils}
  S. G. Nilsson, C. F. Tsang, A. Sobiczewski, Z. Szymanski, 
  S. Wycech, C. Gustafson, I.--L. Lamm, P. M\"oller, B. Nilsson,
  Nucl. Phys. {\bf A131}, 1 (1969).

\bibitem{Dubna}
  Yu. A. Lazarev \emph{et al.},
  Phys. Rev. C {\bf 54}, 620 (1996).

\bibitem{GSI}
  S. Hofmann,
  Rep. Prog. Phys. {\bf 61}, 639 (1998).

\bibitem{GSI2}
  S. Hofmann, G. M\"unzenberg,
  Rev. Mod. Phys. {\bf 72}, 733 (2000).

\bibitem{Z114}
  Yu. Ts. Oganessian \emph{et al.},
  Nature {\bf 400}, 242 (1999),\\
  Yu. Ts. Oganessian \emph{et al.},
  Phys. Rev. Lett. \textbf{83}, 3154 (1999),\\
  Yu. Ts. Oganessian \emph{et al.},
  Phys. Rev. C \textbf{62}, 041604 (2000).

\bibitem{Z116}
  Yu. Ts Oganessian \emph{et al.}, 
  Phys. Rev. C \textbf{63}, 011301(R) (2001).

\bibitem{Z118}
  V. Ninov \emph{et al.},
  Phys. Rev. Lett. {\bf 83}, 1104 (1999).

\bibitem{Arm00a}
  P. Armbruster,
  Eur. Phys. J. \textbf{A7}, 23 (2000). 

\bibitem{Hessberger}
  F. P. Hessberger \emph{et al.},
  Z. Phys. \textbf{A359}, 415 (1997).

\bibitem{Reiter}
  P. Reiter \emph{et al.},
  Phys. Rev. Lett. \textbf{82}, 509 (1999).

\bibitem{Leino}   
  M. Leino \emph{et al.}, 
  Eur. Phys. J. \textbf{A6}, 63 (1999).

\bibitem{gogny}
  J. Decharg{\'e}, D. Gogny,
  Phys. Rev. C \textbf{21}, 1568 (1980).

\bibitem{gogny2}
  J.--F. Berger, J. Decharg{\'e}, D. Gogny,
  Proc. of the ``International Workshop on Nuclear Structure Models'',
  Oak Ridge, Tennessee, March 16--25, 1992,
  edited by R. Bengtsson, J. Draayer, W. Nazarewicz,
  World Scientific, Singapore, 1992.

\bibitem{fayans}
  S. A. Fayans, E. L. Trykov, D. Zawischa,
  Nucl. Phys. \textbf{A568}, 523 (1994),\\
  S. A. Fayans, S. V. Tolokonnikov, E. L. Trykov, D. Zawischa,
  Nucl. Phys. \textbf{A676}, 49 (2000).

\bibitem{madland}
  B. A. Nikolaus, T. Hoch, D. G. Madland, 
  Phys. Rev. C {\bf 46}, 1757 (1992).

\bibitem{BHF}
  W. H. Dickhoff, H. M{\"u}ther, 
  Rep. Prog. Phys. {\bf 11}, 1947 (1992).

\bibitem{HenPan}
  H. Heiselberg, V. Pandharipande, 
  Annu. Rev. Nucl. Part. Sci. \textbf{50}, 481 (2000).

\bibitem{Tmatexp}
  J. W. Negele, D. Vautherin, 
  Phys. Rev. C \textbf{5}, 1472 (1972),\\
  J. W. Negele, D. Vautherin, 
  Phys. Rev. C \textbf{11}, 1031 (1975).

\bibitem{droplet}
  W. D. Myers,
  \emph{Droplet Model of Atomic Nuclei},
  IFI/Plenum, New York, 
  1977.

\bibitem{FRDM}
  P. M\"oller, J. R. Nix, W. D. Myers, W. J. Swiatecki,
  Atom. Nucl. Data Tables \textbf{59}, 185 (1995).

\bibitem{FY}
   P. M\"oller, J. R. Nix,
   Nucl. Phys. {\bf A549}, 84, (1992),\\
   P. M\"oller, J. R. Nix,
   J. Phys. {\bf G 20}, 1681, (1994).

\bibitem{GD90}
  E. K. U. Gross, R. M. Dreizler, 
  \emph{Density functional theory},
  Springer, Berlin 1990.

\bibitem{Petkov}
  I. Zh. Petkov, M. V. Stoitsov,
  \emph{Nuclear Density Functional Theory},
  Clarendon Press, Oxford, 1991.

\bibitem{Dob96a}
  J. Dobaczewski, J. Dudek,
  Proc. of ``High Angular Momentum Phenomena, 
  Workshop in honour of Zdzis{\l}aw Szyma{\'n}ski'', 
  Piaski, Poland, August 23--26, 1995,
  Acta Physica Polonica \textbf{B27}, 45 (1996).

\bibitem{Rei94a}
  P.--G. Reinhard, C. Toepffer, 
  Int. J. Mod. Phys. \textbf{E3}, 435 (1994).

\bibitem{ETFSI}
  Y. Aboussir, J. M. Pearson, A. K. Dutta, F. Tondeur,
  Nucl. Phys. \textbf{A549}, 155 (1992).

\bibitem{TFmyers}
  W. D. Myers, W. J. Swiatecki,  
  Phys. Rev. C \textbf{58}, 3368 (1998),\\
  W. D. Myers, W. J. Swiatecki, 
  Phys. Rev. C \textbf{60}, 54313 (2000).

\bibitem{Skfirst}
  D. Vautherin, D. M. Brink, 
  Phys. Rev. C \textbf{5}, 626 (1972),\\
  M. Beiner, H. Flocard, N. Van Giai, P. Quentin, 
  Nucl. Phys. \textbf{A238}, 29 (1975).

\bibitem{skyrme}
  T. H. R. Skyrme, 
  Phil. Mag. \textbf{1}, 1043 (1956),\\
  T. H. R. Skyrme, 
  Nucl. Phys. \textbf{9}, 615 (1959).

\bibitem{skyrmeLS}
  J. S. Bell, T. H. R. Skyrme, 
  Phil. Mag. \textbf{1}, 1055 (1956),\\
  T. H. R. Skyrme,
  Nucl. Phys. \textbf{9}, 635 (1959).

\bibitem{effmass} 
  M. Jaminon, C. Mahaux, 
  Phys. Rev. {\bf C40}, 354 (1989),
  and references therein.

\pagebreak

\bibitem{lsfirst}
  O. Haxel, J. H. D. Jensen, H. E. Suess,
  Phys. Rev. \textbf{75}, 1766 (1949),\\
  M. G{\"o}ppert--Mayer,
  Phys. Rev. \textbf{75}, 1969 (1949),\\
  M. G{\"o}ppert--Mayer,
  Phys. Rev. \textbf{78}, 16 (1950),\\
  M. G{\"o}ppert--Mayer,
  Phys. Rev. \textbf{78}, 22 (1950).

\bibitem{RingSchuck}
  P. Ring, P. Schuck,
  \emph{The nuclear many-body problem},
  Springer (1980). 

\bibitem{SLyx}
  E. Chabanat, P. Bonche, P. Haensel, J. Meyer, R. Schaeffer,
  Nucl. Phys. \textbf{A635}, 231 (1998),
  Nucl. Phys. \textbf{A643}, 441(E) (1998).

\bibitem{SHF2micmac}
  M. Brack, P. Quentin, 
  Phys. Lett. \textbf{56B}, 421 (1975),\\
  M. Brack, C. Guet, H.--B. H{\aa}kansson, 
  Phys. Rep. \textbf{123}, 275 (1985).

\bibitem{SkIx}
  P.--G. Reinhard, H. Flocard,  
  Nucl. Phys. {\bf A584}, 467 (1995).

\bibitem{RPAnucl}
  P.-G. Reinhard, 
  Ann. Phys. (Leipzig) \textbf{1}, 632 (1992).

\bibitem{duerr}
  H.--P. Duerr,
  Phys. Rev. \textbf{103}, 469 (1955).

\bibitem{RMF1}
  B. D. Serot, J. D. Walecka,
  Phys. Lett. \textbf{87B}, 172 (1979). 

\bibitem{RMF2}
  J. Boguta, A. R. Bodmer,
  Nucl. Phys. \textbf{A292}, 413 (1977).

\bibitem{Schmid}
  R. N. Schmid, E. Engel, R. M. Dreizler,
  Phys. Rev. C \textbf{52}, 164 (1995),\\
  R. N. Schmid, E. Engel, R. M. Dreizler,
  Phys. Rev. C \textbf{52}, 2804 (1995).

\bibitem{Rei89}
  P.--G. Reinhard,
  Rep. Prog. Phys. {\bf 52}, 439 (1989).

\bibitem{Ser92a}
  B. D. Serot, 
  Rep. Prog. Phys. \textbf{55}, 1855 (1992).

\bibitem{Rin96a}
  P. Ring,
  Prog. Part. Nucl. Phys. \textbf{37}, 193 (1996).

\bibitem{PL40}
  P.--G. Reinhard, 
  Z. Phys. \textbf{A329}, 257 (1988).

\bibitem{TM1}
  Y. Sugahara, H. Toki,
  Nucl. Phys. \textbf{A579}, 557 (1994).

\bibitem{thies}
  M. Thies, 
  Phys. Lett. {\bf B166}, 23 (1986)

\bibitem{gappaper}
  M. Bender, K. Rutz, P.--G. Reinhard, J. A. Maruhn,
  Eur. Phys. J. {\bf A8}, 59 (2000).

\bibitem{cmpaper}
  M. Bender, K. Rutz, P.--G. Reinhard, J. A. Maruhn,
  Eur. Phys. J. \textbf{A7}, 467 (2000).

\bibitem{pearstond}
  F. Tondeur, S. Goriely, J. M. Pearson, M. Onsi, 
  Phys. Rev. C \textbf{62}, 024308 (2000).

\bibitem{Fri86a}
  J. Friedrich, P.--G. Reinhard, 
  Phys. Rev. C \textbf{33}, 335 (1986).

\bibitem{SkM*}
   J. Bartel, P. Quentin, M. Brack, C. Guet, H.--B. H{\aa}kansson,
   Nucl. Phys. {\bf A386}, 79 (1982).

\bibitem{SkP}
   J. Dobaczewski, H. Flocard, J. Treiner,
   Nucl. Phys. {\bf A422}, 103 (1984).

\bibitem{Tx}
   F. Tondeur, M. Brack, M. Farine, J. M. Pearson,
   Nucl. Phys. \textbf{A420}, 297 (1984).

\bibitem{NLZ}
   M. Rufa, P.--G. Reinhard, J. A. Maruhn, W. Greiner, M. R. Strayer,
   Phys. Rev. C {\bf 38}, 390 (1989).

\bibitem{NL3}
  G. A. Lalazissis,  J. K\"onig, P. Ring,
  Phys. Rev. C {\bf 55}, 540 (1997).

\bibitem{FriVoe}
  J. Friedrich, N. Voegler, 
  Nucl. Phys. {\bf A373}, 191 (1982).

\bibitem{Ringls}
  M. M. Sharma, G. A. Lalazissis, J. K{\"o}nig, P. Ring,
  Phys. Rev. Lett. \textbf{74}, 3744 (1995).

\bibitem{incomp}
  J. P. Blaizot, J.--F. Berger, J. Decharg{\'e}, M. Girod,
  Nucl. Phys. \textbf{A535}, 331 (1995).

\bibitem{varenna}
  P.--G. Reinhard,
  Nucl. Phys. {\bf A649}, 305c (1999).

\bibitem{Bue98a}
  T. B\"urvenich, K. Rutz, M. Bender, P.--G. Reinhard, J. A. Maruhn, 
  W. Greiner,\\
  Eur. Phys. J. {\bf A3}, 139 (1998).

\bibitem{dubna}
  M. Bender,
  Proc. of the International Conference on 
  ``Fusion dynamics at the extremes'', 
  Dubna, Russia, May 25--27, 2000
  (in print).

\bibitem{RutzDiss}
  K. Rutz,
  Doctoral Dissertation,
  J. W. Goethe Universit\"at Frankfurt am Main,\\
  ibidem--Verlag Stuttgart, 1999.

\bibitem{Ben99a}
  M. Bender, K. Rutz, P.--G. Reinhard, J. A. Maruhn, W. Greiner,
  Phys. Rev. C {\bf 60}, 034304 (1999).

\bibitem{firstSH}
  K. Rutz, M. Bender, T. B\"urvenich, T. Schilling, P.--G. Reinhard, 
  J. A. Maruhn,\\ W. Greiner,
  Phys. Rev. C {\bf 56}, 238 (1997).

\bibitem{shelcor}
  A. T. Kruppa, M. Bender, W. Nazarewicz, P.--G. Reinhard, T. Vertse, 
  S. {\'C}wiok,
  Phys. Rev. C {\bf 61}, 034313 (2000).

\bibitem{newshelcor}
  P.--G. Reinhard, M. Bender, W. Nazarewicz,
  unpublished results.

\bibitem{Rut98a}
  K. Rutz, M. Bender, P.--G. Reinhard, J. A. Maruhn, W. Greiner,
  Nucl. Phys. \textbf{A634}, 67 (1998).

\bibitem{NUDAT} 
  R. R. Kinsey \emph{et al.}, 
  \emph{The NUDAT/PCNUDAT program for nuclear data}, 
  9th International Symposium of Capture Gamma-Ray Spectroscopy and Related Topics
  (Budapest, Hungary), October 1996, Data extracted from the NUDAT database,
  version March 20, 1997, National Nuclear Data Center, Brookhaven National
  Laboratory.

\bibitem{Sb133}
  M. Sanchez--Vega \emph{et al.},
  Phys. Rev. C \textbf{60}, 024303 (1999).

\bibitem{bubble}
  J. Decharg{\'e}, J.-F. Berger, K. Dietrich, M. S. Weiss,
  Phys. Lett. \textbf{B451}, 275 (1999).

\bibitem{shellflu}
  P.--G. Reinhard , J. Friedrich, N. Voegeler, 
  Z. Phys. \textbf{A316}, 207 (1984).

\bibitem{PESfiss}
  M. Bender, K. Rutz, P.--G. Reinhard, J. A. Maruhn, W. Greiner,\\
  Phys. Rev. C {\bf 58}, 2126 (1998).

\bibitem{Cwi96a}
  S. {\'C}wiok, J. Dobaczewski, P.--H. Heenen, P. Magierski, and 
  W. Nazarewicz,\\
  Nucl. Phys. {\bf A611}, 211 (1996).

\bibitem{shapeiso}
  P.--G. Reinhard, D. J. Dean, W. Nazarewicz, J. Dobaczewski, J. A. Maruhn, 
  and \\ M. R. Strayer,
  Phys. Rev. C {\bf 60}, 14316 (1999).

\bibitem{186Pb}
  A. N. Andreyev, {\em et al.},
  Nature {\bf 405}, 430 (2000).

\bibitem{Ber96a}
  J.--F. Berger, L. Bitaud, J. Decharg{\'e}, M. Girod, and S. Peru-Dessenfants, 
  Proc. of the ``International Workshop XXXIV on Gross Properties 
  of Nuclei and Nuclear Exitations'', 
  Hirschegg, Austria, 1996, 
  edited by H. Feldmeier, J. Knoll, and W. N\"orenberg 
  (GSI, Darmstadt, 1996), p. 56. 

\bibitem{Cwi99a}
  S. {\'C}wiok, P.--H. Heenen, W. Nazarewicz, 
  Phys. Rev. Lett. {\bf 83}, 1108 (1999). 

\bibitem{Ben00a}
  M. Bender,
  Phys. Rev. C {\bf 61}, 031302(R) (2000).

\bibitem{Cwi94a} 
  S. {\'C}wiok , S. Hofmann, W. Nazarewicz, 
  Nucl. Phys. {\bf A573}, 356 (1994).

\bibitem{girod}
  M. Girod, P.--G. Reinhard,
  Nucl. Phys. \textbf{A384}, 179 (1982).

\bibitem{Rei00a}
  P.--G. Reinhard, M. Bender, T. B\"urvenich, T. Cornelius, P. Fleischer, 
  J. A. Maruhn,
  Proc. of the ``Tours Symposium on Nuclear Physics IV'', 
  Tours, France, September 2000
  AIP, in print. 

\end{thebibliography}
\end{document}